\documentclass[final,5p,twocolumn,number]{elsarticle} 
\usepackage{multicol}

\usepackage{amsmath}
\usepackage{amsfonts}
\usepackage{amssymb}
\usepackage{amsthm}
\usepackage{mathtools}
\usepackage{physics}
\usepackage{siunitx}
\usepackage{optidef}
\usepackage{eurosym}

\usepackage{adjustbox}
\usepackage{epstopdf}
\usepackage{float}
\usepackage{graphicx}
\usepackage{rotating}
\usepackage{subcaption}

\usepackage{tikz}
\usepackage{pgfplots}
\usepackage{pgfplotstable}
\pgfplotsset{compat=1.10}
\usepgfplotslibrary{fillbetween}
\usetikzlibrary{calc}
\usetikzlibrary{positioning}
\usetikzlibrary{decorations.markings}
\usetikzlibrary{trees}
\usetikzlibrary{patterns}
\usetikzlibrary{decorations.pathreplacing}
\usetikzlibrary{backgrounds,patterns,calc}
\usepackage{tikz-qtree}

\usepackage{booktabs}
\usepackage{ctable}
\usepackage{longtable}
\usepackage{threeparttable}
\usepackage{tabularx}
\usepackage{multirow}

\usepackage{xcolor}
\usepackage{color}
\usepackage[ruled]{algorithm2e}
\usepackage[labelfont={bf}]{caption}
\usepackage{enumerate}
\usepackage{lastpage}
\usepackage{listings}
\usepackage[hyphens]{url}
\usepackage[colorlinks=true, allcolors=blue]{hyperref}
\usepackage[title]{appendix}

\usepackage{verbatim}

\date{\today} 
\allowdisplaybreaks[1]

\theoremstyle{definition}

\journal{arXiv}

\begin{document}

\begin{frontmatter}

\title{Economic incentives for capacity reductions on interconnectors in the day-ahead market}


\author[NTNU]{E. Ruben van Beesten\corref{correspondingauthor}}
\ead{ruben.van.beesten@ntnu.no}
\cortext[correspondingauthor]{Corresponding author}

\author[RuG]{Daan Hulshof}
\ead{d.hulshof@rug.nl}

\address[NTNU]{Department of Industrial Economics and Technology Management, Norwegian University of Science and Technology (NTNU), Alfred Getz' vei 1, 7034, Trondheim, Norway}

\address[RuG]{Faculty of Economics and Business, University of Groningen, Nettelbosje 2, 9747 AE, Groningen, the Netherlands}

\begin{abstract}
    We consider a zonal international power market and investigate potential economic incentives for short-term reductions of transmission capacities on existing interconnectors by the responsible transmission system operators (TSOs). We show that if a TSO aims to maximize domestic total welfare, it often has an incentive to reduce the capacity on the interconnectors to neighboring countries.
    In contrast with the (limited) literature on this subject, which focuses on incentives through the avoidance of future balancing costs, we show that incentives can exist even if one ignores balancing and focuses solely on welfare gains in the day-ahead market itself. Our analysis consists of two parts. In the first part, we develop an analytical framework that explains why these incentives exist. In particular, we distinguish two mechanisms: one based on price differences with neighboring countries and one based on the domestic electricity price. In the second part, we perform numerical experiments using a model of the Northern-European power system, focusing on the Danish TSO. In 97\% of the historical hours tested, we indeed observe economic incentives for capacity reductions, leading to significant welfare gains for Denmark and welfare losses for the system as a whole. We show that the potential for welfare gains greatly depends on the ability of the TSO to adapt interconnector capacities to short-term market conditions. Finally, we explore the extent to which the recently introduced European ``70\%-rule'' can mitigate the incentives for capacity reductions and their welfare effects.
\end{abstract}


\end{frontmatter}

\section{Introduction}
\label{sec:introduction}

Transmission cables connecting the electricity grids of different countries are a vital element in the electricity system, enabling international trade in electricity to happen. Due to differences between regions, mainly in supply conditions associated with differences in, e.g., environmental conditions and weather patterns, electricity may be abundantly available in one country while at the same time being scarcely available in another country. Transmission cables between different countries, referred to as \textit{interconnectors}, enable electricity to flow from the abundant country to the scarce country, resulting in the typical welfare gains associated with international trade \cite{beugelsdijk2013international}. In the coming decades, the importance of this international trade will likely even increase, given the electrification trends (e.g., of transport and buildings) as well as the increasing variability of electricity supply associated with the transition to renewable sources like wind and solar \cite{moradi2021integrated}. The latter means that the differences in scarcity between regions will further increase, tending to increase the welfare gains from trade and, hence, the economic importance of interconnectors.


To realize these welfare gains, interconnectors need to be utilized efficiently. That is, its physical transmission capacity should be made available for trading as much as possible. However, there are several technical reasons why full availability of an interconnector's capacity is not always possible, including bottlenecks in the domestic grids of different countries and unscheduled ``loop'' flows that enter the regional grid from another region. To account for these technical restrictions and safeguard the reliability of the grid, European regulation assigns Transmission System Operators (TSOs) with the responsibility to impose limits on the transmission capacity made available for trading on the interconnectors their transmission grids are connected to \cite{horn2021national_new}. The regulations specify the precise conditions under which capacity restrictions on interonnectors can be imposed (see \cite{horn2021national_new} for an overview of the existing regulations in the European power market).

However, the ability to restrict transmission capacities on interconnectors clearly also gives TSOs market power: their decisions can affect the outcome of the electricity markets. This market power could potentially be abused by a TSO to increase its profits. First, it could be used to increase congestion rents earned on the interconnectors it is connected to (typically, congestion rents are split 50/50 between the two TSOs operating the grid on both ends of an interconnector). Second, it could be used to avoid future balancing costs for the TSO \cite{glachant2005nordic}. Besides these direct TSO profits, TSOs could potentially also abuse their market power to increase \textit{domestic welfare}, i.e., the national welfare of the country in which they operate the grid \cite{horn2021national_new}. Incentives to do so may arise from the fact that European TSOs are typically 100\% owned by their respective national government.

The fact that this potential for market power abuse exists in theory does not directly imply that it also happens in practice. 
However, there are some empirical signals pointing in this direction.
For instance, in 2018 two Danish wind energy lobby groups filed a complaint about the Swedish TSO, who they alleged ``discriminates between internal and cross border flows of electricity'' and ``distorts economic signals to both market participants and TSOs'' by its capacity restrictions on the two Danish-Swedish interconnectors \cite{winddenmark2018}. %
Similarly, in November 2021 the Norwegian TSO pointed out that the Swedish TSO was applying export restrictions on the cable connecting the southern regions of the two countries. Both the Norwegian TSO and regulators appeared to question to some degree whether these restrictions are strictly necessary for technical reasons \cite{Statnett2021letter, e24NorwayConcerned}, while their economic consequences were substantial, given the high electricity prices in Europe at the time. %

We do not aim to assess the validity of these specific complaints. Instead, this paper aims to contribute to the theoretical understanding of economic incentives for capacity reductions on interconnectors and the welfare effects of these reductions. We are particularly interested in the potential of increasing domestic welfare through capacity restrictions.

In the literature, there are several lines of research that study long-term decisions on capacities of interconnectors and the resulting economic benefits.
For instance, the literature on market coupling focuses on the question whether and how to couple different electricity markets together in an international system. Some papers explicitly investigate the welfare effects of market coupling \cite{hobbs2005more,pellini2012measuring,OCHOA2015522}, while others show that the effective transmission capacity can be increased by accounting for the physical reality of electricity flows in sophisticated ways, such as through flow-based market coupling \cite{van2016flow, kristiansen2020flow}. 
Also related is the literature on transmission expansion planing, aiming to answer the question of where to build new (international) transmission cables in order to increase system welfare \cite{lumbreras2016new, mahdavi2018transmission, gomes2019state}. Some papers also investigate the question of how to distribute the welfare gains from proposed transmission expansion plans \cite{churkin2021review}. 
All these strands of literature have one thing in common: they focus on \textit{long-term} decisions on international transmission capacity, whether in the form of market coupling or transmission expansion. However, once constructed, TSOs have the ability to influence the capacity on an interconnector \textit{intermittently}: they can impose different capacity restrictions for every hour in the day-ahead market. 

To our knowledge, only two studies explicitly investigate the potential for abusing the market power resulting from the ability to impose intermittent capacity restrictions on existing interconnectors. First, Glachant and Pignon \cite{glachant2005nordic} use a stylized model of the Nordic electricity market to show that intermittent capacity restrictions on interconnectors in the day-ahead market can increase the congestion rent earned by TSOs, while reducing the domestic balancing costs. Second, Horn and Tanger{\aa}s \cite{horn2021national_new} develop a theoretical framework to analyze the intermittent incentives of TSOs to reduce transmission capacity on existing interconnectors under the assumption that TSOs maximize domestic welfare (this assumption is justified by the fact that TSOs are typically state-owned organizations and thus can be expected to act in the interest of its host nation). Their analysis also focuses on the role of the balancing market and includes a proposal for a new balancing market design that eliminates the identified economic incentives for intermittent capacity restrictions.

This paper adds to this scarce strand of literature and investigates to what extent economic incentives for imposing intermittent capacity restrictions on interconnectors may exist in the day-ahead market. While the literature focuses mainly on the avoidance of balancing costs in the balancing market, we focus purely on the day-ahead market itself. We show that, even when balancing costs are ignored, economic incentives for reducing interconnector capacities can exist in the form of an increase in domestic welfare generated in the day-ahead market.
We explicitly outline the economic mechanisms causing these domestic welfare gains and we analyze the impact of nationally optimal capacity reductions on the distribution of welfare over domestic and foreign consumers, producers and TSOs. Another novelty of our paper is that we study these restrictions numerically, using a model of the Northern-European power system that is based on historical data for consumption, production, and price levels, as well as actual interconnection capacities between price zones. This allows us to study the magnitude of the welfare effects resulting from capacity reductions, as well as the effectiveness of regulations in mitigating adverse welfare effects on other countries.

Our analysis consists of two parts. In the first part we provide an analytical framework that illustrates that domestic welfare-maximizing TSOs may have economic incentives to reduce transmission capacity on interconnectors. Using simple illustrative examples, we identify two mechanisms through which these economic incentives can arise: the \textit{price difference mechanism}, based on creating price differences with foreign price zones and inducing congestion rent, and the \textit{domestic price mechanism}, based on changing the domestic price and increasing the sum of producer and consumer surplus. 

In the second part we perform numerical experiments using a model of the Northern-European power system. Based on historical data, this power system includes the supply and demand and cross-border transmission grid characteristics of eleven countries that form a meshed electricity network in practice. In the experiments we focus on investigating the incentives for reducing transmission capacity by the Danish TSO, who occupies a central position in the geographical area we consider. We show that in most historical hours tested, incentives for capacity reductions indeed exist and we quantify the resulting welfare effects. Comparing these findings with a setting where only a single long-term capacity reduction is possible, we highlight how our results contrast with the literature on, e.g., market coupling and transmission expansion planning. Finally, we compare our results with a setting where the recently introduced European ``70\% rule'' \cite{ACER2022report}, stating that at least 70\% of each interconnection should be available at all times, is enforced.\footnote{The 70\% rule was implemented in 2020 but TSOs can apply for an exemption. In practice, the rule is frequently violated, particularly in continental Europe, but also on interconnectors between Denmark and Sweden, and Finland and Sweden, for instance \cite{ACER2022report}.} We show that enforcement of this rule does not remove economic incentives for capacity reductions, but does mitigate the associated welfare effects to some extent.

The remainder of this paper is organized as follows. Section \ref{sec:analytical_framework} provides the analytical framework and discusses mechanisms through which TSOs may increase domestic welfare by capacity restrictions. Section \ref{sec:case_study} describes the model of the representative power system as well as the experimental design to analyze the presence and magnitude of the welfare impact of TSO restrictions. The results are discussed in Section \ref{sec:results}. Section \ref{sec:conclusion} concludes the paper. Finally, Appendices~\ref{sec:mathematical_formulation} and \ref{sec:data} provide a detailed description of the mathematical model and the data used in the case study, respectively.

\section{Analytical framework} 
\label{sec:analytical_framework}

This section presents two stylized examples of electricity trade between countries in order to illustrate how restrictions on interconnectors can manipulate prices and increase domestic welfare in one country.\footnote{Throughout this section, we assume that a country has a single TSO and that this TSO aims to maximize the domestic welfare. This resembles the situation in most European countries, except for Germany and Austria, which have multiple TSOs. Furthermore, when referring to welfare, we refer to welfare in the power sector itself only. Indirect welfare effects in other sectors are ignored.} The two examples describe distinct mechanisms through which restrictions increase domestic welfare: the \textit{price difference mechanism}, based on creating price differences with neighboring countries, and the \textit{domestic price mechanism}, based on changing the domestic price. In Section \ref{subsec:2node} we illustrate the price difference mechanism. We use a simple two-zone setting to first show the standard result that trade increases welfare in the system as a whole. We then show that capacity restrictions can lead to welfare gains by creating price differences with neighbors that induce congestion rent. Crucial in this mechanism is the assumption that TSOs share congestion rent on interconnectors equally between each other, as is common practice in the European Union (EU) \cite{huang2016mind}. In Section~\ref{subsec:3node} we illustrate the domestic price mechanism. 
Using a three-zone setting, we show that a TSO can use capacity restrictions on interconnectors in order to block access of consumers in neighboring import zones or producers in neighboring export zones to the electricity traded domestically. This affects the domestic scarcity of electricity, reflected in the domestic price. In turn, this affects the domestic producer and consumer surplus and can increase domestic welfare.

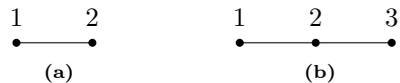
\begin{figure}[h]
    \centering
    \begin{subfigure}[b]{0.25\columnwidth}
        \centering
        \begin{tikzpicture}[scale=0.5]
    \coordinate (n1) at (0,0);
    \coordinate (n2) at (2,0);
    \draw (n1) -- (n2);
    \fill[black] (n1) circle (3pt);
    \fill[black] (n2) circle (3pt);
    \node[above=1mm of n1] {1};
    \node[above=1mm of n2] {2};
\end{tikzpicture}
        \caption{}
        \label{subfig:2node_config}
    \end{subfigure}
    \begin{subfigure}[b]{0.5\columnwidth}
        \centering
        \begin{tikzpicture}[scale=0.5]
    \coordinate (n1) at (0,0);
    \coordinate (n2) at (2,0);
    \coordinate (n3) at (4,0);
    \draw (n1) -- (n2);
    \draw (n2) -- (n3);
    \fill[black] (n1) circle (3pt);
    \fill[black] (n2) circle (3pt);
    \fill[black] (n3) circle (3pt);
    \node[above=1mm of n1] {1};
    \node[above=1mm of n2] {2};
    \node[above=1mm of n3] {3};
\end{tikzpicture}
        \caption{}
        \label{subfig:3node_config}
    \end{subfigure}
    \caption{Schematic illustration of two-zone and three-zone electricity systems.}
    \label{fig:network_configurations}
\end{figure}

\subsection{Price difference mechanism} 
\label{subsec:2node}

\begin{figure}[t]
    \centering
    \begin{tikzpicture}[scale=0.6]
    \draw [<->] (0,11) node (yaxis) [above] {$p$} |- (5,0) node (xaxis) [right] {$q$};
    \draw[domain=0:5,smooth,variable=\x,blue] plot ({\x},{10-\x});
    \coordinate (D1) at (5,5);
    \node[right=0mm of D1]{$D_1$};
    \coordinate (dchoke) at (0,10);
    \draw[domain=0:4,smooth,variable=\x,red] plot ({\x},{2+2*\x});
    \coordinate (S1) at (4,10);
    \node[right=0mm of S1]{$S_1$};
    \coordinate (schoke) at (0,2);
    \coordinate (uncap) at (2.67,7.33);
    \draw[dashed] (yaxis |- uncap) node[left] {$p_1^*$} -| (xaxis -| uncap) node[below] {$q_1^*$};
    \coordinate (capdem) at (3.68,6.32);
    \coordinate (capsup) at (2.16,6.32);
    \coordinate (p2) at (3.68,5.02);
    \draw[dashed] (yaxis |- capsup) node[left] {$p_1'$} -| (xaxis -| capsup);
    \draw[dashed] (capsup|- capdem) node[left] {} -| (xaxis -| capdem);
    \coordinate (optdem) at (4.33,5.67);
    \coordinate (optsup) at (1.835,5.67);
    \draw[dashed] (yaxis |- optsup) node[left] {$\bar{p}$} -| (xaxis -| optsup) node[left, anchor=north] {$s_1$};
    \draw[dashed] (yaxis |- optdem) node[left] {} -| (xaxis -| optdem) node[below] {$d_1$};
    \fill[orange,opacity=0.2] plot coordinates {(0,10)(0,5.67)(optsup)(uncap)};
    \fill[green,opacity=0.2] plot coordinates {(optsup)(uncap)(optdem)};
    
    \fill[cyan,opacity=0.2] plot coordinates {(0,5.67)(1.835,5.67)(0,2)};
\end{tikzpicture}
\quad
\begin{tikzpicture}[scale=0.6]
    \draw [<->] (0,11) node (yaxis) [above] {$p$} |- (6,0) node (xaxis) [right] {$q$};
    \draw[domain=0:5,smooth,variable=\x,blue] plot ({\x},{10-2*\x});
    \coordinate (D1) at (1,8);
    \node[right=0mm of D1]{$D_2$};
    \draw[domain=0:6,smooth,variable=\x,red] plot({\x},{1+\x});
    \coordinate (S1) at (6,7);
    \node[right=0mm of S1]{$S_2$};
    \coordinate (uncap) at (3,4);
    \draw[dashed] (yaxis |- uncap) node[left] {$p_2^*$} -| (xaxis -| uncap) node[below] {$q_2^*$};
    \coordinate (capdem) at (2.49,5.02);
    \coordinate (capsup) at (4.02,5.02);
    \draw[dashed] (yaxis |- capdem) node[left] {$p_2^\prime$} -| (xaxis -| capdem);
    \draw[dashed] (capdem |- capsup) node[left] {} -| (xaxis -| capsup);
    \coordinate (optdem) at (2.165,5.67);
    \coordinate (optsup) at (4.67,5.67);
    \draw[dashed] (yaxis |- optsup) node[left] {$\bar{p}$} -| (xaxis -| optsup) node[below] {$s_2$};
    \draw[dashed] (yaxis |- optdem) node[left] {} -| (xaxis -| optdem) node[below] {$d_2$};
   
    \fill[orange,opacity=0.2] plot coordinates {(0,10)(0,5.67)(optdem)};
    
    
    \fill[cyan,opacity=0.2] plot coordinates {(0,5.67)(0,1)(uncap)(optdem)};
    
    \fill[yellow,opacity=0.25] plot coordinates {(optsup)(uncap)(optdem)};
\end{tikzpicture}
    \caption{Supply and demand curves in zone 1 (top) and zone 2 (bottom) in the two-zone example. The orange and green areas represent CS; the blue and yellow areas represent PS.}
    \label{fig:surplus_two_nodes}
\end{figure}
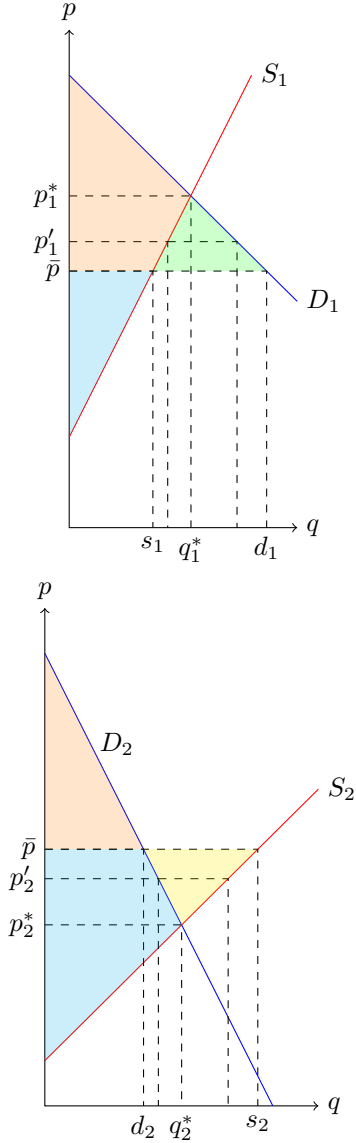

Consider a simple setting with two zones, each of which represents a distinct electricity grid with its own electricity producers and electricity consumers (see Figure~\ref{subfig:2node_config} for an illustration). Producers and consumers are  represented by the linear supply and demand curves $S_i(p_i)$ and $D_i(p_i)$, respectively, where $S_i$ and $D_i$ are the quantity $q$ of supply and demand, respectively, and $p_i$ is the price in zone $i=1,2$. Figure~\ref{fig:surplus_two_nodes} displays the supply and demand curves in the two zones. We assume that the market is characterized by perfect competition.

Without inter-zonal trade, the equilibrium market outcome is determined by the point of intersection of the domestic supply and demand curves. Define the equilibrium price and quantity for zone $i$ by $p_i=p_i^*$ and $q_1=q_i^*$, respectively. In the example of Figure \ref{fig:surplus_two_nodes} we have $p_1^*>p_2^*$. This price difference implies that producers in zone 2 would be willing to supply electricity to zone 1 and consumers in zone 1 would be willing buy from them, provided this was possible. 

\begin{figure}[h]
    \centering
    \begin{tikzpicture}[scale=0.6]
    \coordinate (o) at (0,0);
    \coordinate (xmax) at (4,0);
    \coordinate (ymax) at (0,8);
    \draw[->] (o) -- (xmax);
    \node[right=0mm of xmax]{$q$};
    \draw[->] (o) -- (ymax);
    \node[above=0mm of ymax]{$p$};
    
    \draw[domain=0:4,smooth,variable=\x,blue] plot ({\x},{7.33-0.67*\x});
    \coordinate (p1star) at (0,7.33);
    \node[left=0mm of p1star]{$p_1^*$};
    \coordinate (I1) at (4,4.6);
    \node[right=0mm of I1]{$I_1$};
    
    \draw[domain=0:4,smooth,variable=\x,red] plot ({\x},{0.67*\x+4});
    \coordinate (p2star) at (0,4);
    \node[left=0mm of p2star]{$p_2^*$};
    \coordinate (E2) at (4,6.7);
    \node[right=0mm of E2]{$E_2$};
    
    \coordinate (kopt) at (1.53,0);
    \coordinate (kkopt) at (1.53,6.32);
    \draw[dashed] (kopt) -- (kkopt);
    \node[below=0mm of kopt]{$q'$};
    
    \coordinate (p1) at (0,6.32);
    \coordinate (p2) at (0,5.02);
    \coordinate (opt1) at (1.53,6.32);
    \coordinate (opt2) at (1.53,5.02);
    \coordinate (pnew) at (0,5.67);
    \coordinate (optnew) at (2.5, 5.67);
    \draw[dashed] (p1) -- (opt1);
    \node[left=0mm of p1]{$p_1'$};
    \draw[dashed] (p2) -- (opt2);
    \node[left=0mm of p2]{$p_2'$};
    \draw[dashed] (pnew) -- (optnew);
    \node[left=1mm of pnew]{$\bar{p}$};
    
    \fill[green,opacity=0.2] plot coordinates {(0,7.33)(optnew)(pnew)};
    
    \fill[yellow,opacity=0.25] plot coordinates {(0,4)(optnew)(pnew)};

    
    \coordinate (opt) at (2.5,5.67);
    \coordinate (kmax) at (2.5,0);
    \draw[dashed] (kmax) -- (opt);
    \node[below=0mm of kmax]{$\bar{q}$};
    
    
    \fill[pattern=north east lines, pattern color=black,opacity=1]  plot coordinates {(p1)(opt1)(1.53,5.67)(pnew)};
    \fill[pattern=north east lines, pattern color=black,opacity=1]  plot coordinates {(p2)(opt2)(1.53,5.67)(pnew)};
    
    \fill[pattern=crosshatch dots, pattern color=black,opacity=1]  plot coordinates {(opt1)(1.53,5.67)(opt)};
    \fill[pattern=crosshatch dots, pattern color=black,opacity=1]  plot coordinates {(opt2)(1.53,5.67)(opt)};

\end{tikzpicture}
    \caption{Import/Export graph for the two-zone example. The green and yellow areas represent the welfare gain from unrestricted trade in zone 1 and 2, respectively. A capacity restriction to $q'$ decreases producer and consumer surplus by the striped and dotted areas and increases congestion rent by the striped areas. System welfare decreases by the dotted areas.}
    \label{fig:2_node_trade_fig}
\end{figure}
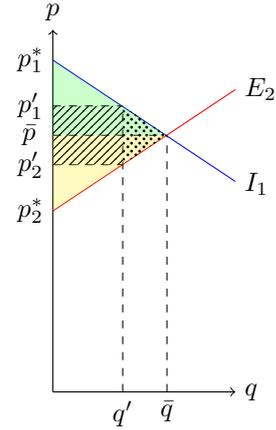

An interconnector between the two zones enables trade between suppliers and consumers in the different regions. The willingness of consumers in zone 1 to import from zone 2 can be represented by the import curve $I_1=D_1-S_1$. For each price, this curve provides the quantity that consumers in zone 1 are willing to consume in excess of what domestic producers are willing to supply, i.e., the willingness to import. In an equivalent fashion, the willingness of producers in zone 2 to export to zone 1 can be represented by the export curve $E_2=S_2-D_2$. Figure~\ref{fig:2_node_trade_fig} displays these import and export curves. In line with the common practice on interconnectors in the EU, we assume here that the direct financial benefit associated with electricity trade, i.e., the congestion rent $CR$, is divided equally among the two TSOs that operate the grids in the connected zones. $CR$ is equal to the difference in the market price between two connected zones, multiplied by the quantity exported through the cable.

The market equilibrium with unrestricted trade -- i.e., assuming that an interconnector with unlimited capacity connects the two zones -- is given by the point of intersection of the import and export curves, i.e., $I_1=E_2$, yielding a price and trade quantity of $\bar{p}$ and $\bar{q}$, respectively. Unrestricted trade implies full market integration with $p_1=p_2=\bar{p}$, for otherwise arbitrage opportunities would remain. In turn, identical prices in the two zones implies that $CR=0$ with unrestricted trade.

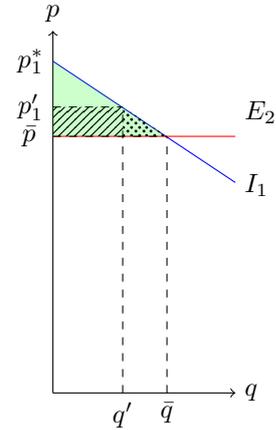
\begin{figure}[h]
    \centering
    \begin{tikzpicture}[scale=0.6]
    \coordinate (o) at (0,0);
    \coordinate (xmax) at (4,0);
    \coordinate (ymax) at (0,8);
    \draw[->] (o) -- (xmax);
    \node[right=0mm of xmax]{$q$};
    \draw[->] (o) -- (ymax);
    \node[above=0mm of ymax]{$p$};
    
    \draw[domain=0:4,smooth,variable=\x,blue] plot ({\x},{7.33-0.67*\x});
    \coordinate (p1star) at (0,7.33);
    \node[left=0mm of p1star]{$p_1^*$};
    \coordinate (I1) at (4,4.6);
    \node[right=0mm of I1]{$I_1$};
    
    \draw[domain=0:4,smooth,variable=\x,red] plot ({\x},{5.67});
    \coordinate (E2) at (4,6.2);
    \node[right=0mm of E2]{$E_2$};
    
    \coordinate (kopt) at (1.53,0);
    \draw[dashed] (kopt) -- (1.53,6.32);
    \node[below=0mm of kopt]{$q'$};
    
    \coordinate (p1) at (0,6.32);
    \coordinate (opt1) at (1.53,6.32);
    \coordinate (pnew) at (0,5.67);
    \coordinate (optnew) at (2.5, 5.67);
    \draw[dashed] (p1) -- (opt1);
    \node[left=0mm of p1]{$p_1'$};
    
    \draw[dashed] (pnew) -- (optnew);
    \node[left=1mm of pnew]{$\bar{p}$};
    
    \fill[green,opacity=0.2] plot coordinates {(0,7.33)(optnew)(pnew)};
    
    
    \coordinate (opt) at (2.5,5.67);
    \coordinate (kmax) at (2.5,0);
    \draw[dashed] (kmax) -- (opt);
    \node[below=0mm of kmax]{$\bar{q}$};
    
    
    \fill[pattern=north east lines, pattern color=black,opacity=1]  plot coordinates {(p1)(opt1)(1.53,5.67)(pnew)};
    
    \fill[pattern=crosshatch dots, pattern color=black,opacity=1]  plot coordinates {(opt1)(1.53,5.67)(opt)};
    
\end{tikzpicture}
    \caption{Import/Export graph illustrating the price difference mechanism in the two-zone example with a horizontal export curve.}
    \label{fig:2_node_trade_fig2}
\end{figure}

Trade increases total welfare in the combined electricity system, as can be illustrated with Figures~\ref{fig:surplus_two_nodes} and \ref{fig:2_node_trade_fig}. As per convention, define consumer surplus ($CS$) as the difference between consumer willingness to pay and the market price aggregated over consumption (i.e., the area below the demand curve and above the market price in Figure \ref{fig:surplus_two_nodes}); define producer surplus ($PS$) as the the difference between the minimally-required producer price and the market price aggregated over production (i.e., the area above the supply curve and below the market price); and define total welfare ($TW$) in zone $i$ as $TW_i=PS_i+CS_i+CR_i$. In zone 1, unrestricted trade decreases the domestic market price from $p^*_1$ to $\bar{p}$, which increases consumer surplus and decreases producer surplus. The increase in $CS_1$ is greater than the decrease in $PS_1$ because the decrease in the latter fully transfers to the former, while $CS_1$ rises beyond that. This `excess' increase in $CS_1$ is represented by the green areas in Figure~\ref{fig:surplus_two_nodes} and equals the increase in $TW_1$, i.e., the net benefit from trade in zone 1. In Figure \ref{fig:2_node_trade_fig}, the increase in $TW_1$ is equivalently represented by the area below the import curve and above the market price, i.e., a green triangle with the same area. Notice that, from the perspective of the importing region, a lower price increases domestic welfare.

The welfare effects in zone 2 are analogous, but in the opposite direction. Here, the domestic market price increases such that $PS_2$ increases and $CS_2$ decreases. The increase in $PS_2$ outweighs the decrease in $CS_2$, thereby increasing $TW_2$. This welfare increase is represented by the yellow areas in Figures~\ref{fig:surplus_two_nodes} and \ref{fig:2_node_trade_fig}. In contrast with the importing region, in an export region, a higher price increases domestic welfare. From the perspective of the electricity system, with both $TW_1$ and $TW_2$ increasing, aggregate system welfare increases as a result of trade through the interconnector. 

If the capacity on the interconnector is reduced to a value below $\bar{q}$, aggregate system welfare is unequivocally reduced. This can be most easily illustrated with Figure \ref{fig:2_node_trade_fig}. Suppose that one of the TSOs decides to limit the capacity on the interconnector, say to $q'$. The interconnector capacity then becomes a binding constraint and prices cannot equalize between the two zones, implying that domestic prices $p_1'$ and $p_2'$ emerge. This reduces the sum of $CS$ and $PS$ in both zones. In Figure \ref{fig:2_node_trade_fig}, for zone 1 (zone 2), the reduction equals the striped and dotted green (yellow) areas. Part of this reduction in welfare, however, is `recouped' in the form of an increase in congestion rent from zero to $(p_1'-p_2')q'$, which is represented by the striped areas. Both TSOs obtain half of this. The reduction in system welfare, or deadweight loss, is equal to the dotted areas.

In the current case with symmetrical import and export curves, for both zones, the increase in the congestion rent that they obtain (the striped green (yellow) area for zone 1 (2)) is smaller than the decrease in the sum of domestic consumer and producer surplus (the dotted green (yellow) area for zone 1 (2)). This means that domestic welfare decreases in both zones and both TSOs can thus not increase domestic welfare by reducing capacity on the interconnector. This last point, however, is not always true. 

To illustrate that domestic welfare can increase from a capacity reduction on the interconnector, imagine that $E_2$ is a flat horizontal curve at level $\bar{p}$. This corresponds to a supply curve in zone 2 with constant marginal costs and abundant production capacity.\footnote{Alternatively, the domestic price mechanism could also be illustrated with a flat horizontal import curve, which would imply that consumers are willing to buy any amount of electricity when the market price is at least equal to a certain threshold.} Figure \ref{fig:2_node_trade_fig2} shows the import and export curves in this new situation. Without restrictions, the efficient level of trade is still given by $\bar{y}$ with an integrated price level of $\bar{p}$. Note that, in this case, trade over the interconnector improves welfare in zone 1 but leaves it unaffected in zone 2.

Consider the same capacity restriction as before on the interconnector, to level $q'$. This restriction results in a price increase in zone 1 to $p_1'$, while the price in zone 2 remains $\bar{p}$. Hence, while $CS_2+PS_2$ remains unaffected, $CS_1+PS_1$ is reduced due to the constraint. In Figure \ref{fig:2_node_trade_fig2}, this reduction is again equal to the striped and dotted green areas. Part of this reduction in welfare is again compensated by an increase in aggregate $CR$, from zero to $(p_1'-\bar{p})q'$, which is represented by the striped green area. In this case, both TSOs obtain half of this, meaning that total welfare decreases in zone 1, as well as in the aggregate system, but \textit{increases} in zone 2. Thus, the TSO in zone 2 can increase its domestic welfare by inducing a price difference with node 1 through a capacity reduction. We refer to this as the \textit{price difference mechanism}.

We point out that the price difference mechanism only increases total domestic welfare if the increase in domestically earned congestion rent exceeds the decrease in the sum of domestic consumer and producer surplus. This decrease equals zero in our simplistic example, but in practice depends on the supply and demand conditions and the resulting import and export curves. In principle, reducing capacity by a TSO will tend to increase domestic welfare as long as the change in the domestic price is relatively small (implying small changes in domestic $PS$ and $CS$) and the change in the foreign price is relatively large (inducing a large price difference and thus high $CR$). Finally, we note that the existence of the price difference mechanism is a direct consequence of the fact that congestion rents on interconnectors are shared between the two TSOs on each end of the cable.


\subsection{Domestic price mechanism} 
\label{subsec:3node}

This section illustrates another mechanism through which TSOs may increase domestic welfare, based on manipulating the domestic price rather than price differences with neighbors. To illustrate this \textit{domestic price mechanism}, we need to consider a system with at least three zones. We consider the three-zone system illustrated in Figure~\ref{subfig:3node_config}. To facilitate the interpretation of the domestic price mechanism, we make a simplifying assumption. In particular, we assume that zone 1 is a pure export region, meaning it only has supply but no demand, while zone 2 and 3 are pure import regions, meaning they have demand but no supply. %
This assumption allows us to interpret total welfare effects in the export zone purely in terms of producer surplus and those in the import zones purely in terms of consumer surplus. However, also without this assumption the arguments laid out below remain valid.

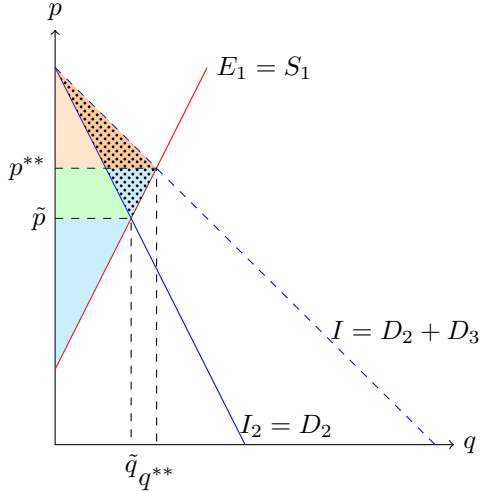
\begin{figure}[t]
    \centering
    \begin{tikzpicture}[scale=0.5]
    \draw [<->] (0,11) node (yaxis) [above] {$p$} |- (10.5,0) node (xaxis) [right] {$q$};
    
    \draw[domain=0:4,smooth,variable=\x,red] plot ({\x},{2+2*\x});
    \coordinate (S0E0) at (4,10);
    \node[right=0mm of S0E0]{$E_1=S_1$};
    
    \draw[domain=0:5,smooth,variable=\x,blue] plot ({\x},{10-2*\x});
    \coordinate (D1E1) at (4.6,0.5);
    \node[right=0mm of D1E1]{$I_2=D_2$};
    
    \draw[domain=0:10,dashed,variable=\x,blue] plot ({\x},{10-\x});
    \coordinate (D12) at (7,3);
    \node[right=0mm of D12]{$I=D_2+D_3$};
    
    \coordinate (pqnolimit) at (2.67,7.33);
    \draw[dashed] (yaxis |- pqnolimit) node[left] {$p^{**}$} -| (xaxis -| pqnolimit) node[below=1.75mm] {$q^{**}$};
    \coordinate (pqlimit) at (2,6);
    \draw[dashed] (yaxis |- pqlimit) node[left] {$\tilde{p}$} -| (xaxis -| pqlimit) node[below] {$\tilde{q}$};
    
    \fill[orange,opacity=0.2] plot coordinates {(0,7.33)(1.33,7.33)(0,10)};
    \fill[green,opacity=0.2] plot coordinates {(0,6)(2,6)(1.33,7.33)(0,7.33)};
     
    \fill[cyan,opacity=0.2] plot coordinates {(1.33,7.33)(2,6)(2.67,7.33)};
    \fill[pattern=crosshatch dots, pattern color=black,opacity=1]  plot coordinates {(1.33,7.33)(2,6)(2.67,7.33)};
    \fill[orange,opacity=0.4] plot coordinates {(0,10)(1.33,7.33)(2.67,7.33)};
    \fill[pattern=crosshatch dots, pattern color=black,opacity=1]  plot coordinates {(0,10)(1.33,7.33)(2.67,7.33)};
    
    \fill[cyan,opacity=0.2] plot coordinates {(0,2)(2,6)(0,6)};
    
\end{tikzpicture}
    \caption{Export/Import graph illustrating the domestic price mechanism in the three-zone example. The green area represents the increase in consumer surplus and welfare of zone 2 due to the capacity restriction. The dotted areas represent the system welfare loss.
}
    \label{fig:3_node_trade_fig}
\end{figure}

In the initial situation with unrestricted trade, electricity can flow freely between regions. The three regions essentially form an integrated market with a single price ($p=p_1=p_2=p_3$), where supply is given by $S_1$, the supply curve in zone 1, and total demand by $D=D_2+D_3$, the joint demand curves of zones 2 and 3. Figure \ref{fig:3_node_trade_fig} depicts these supply and demand curves in a single figure. In this simplistic case, the export curve in zone 1 is equal to the supply curve (i.e., $E_1 = S_1$) and the import curves in zone 2 and 3 are equal to the respective demand curves (i.e., $I_3 = D_3=I_2=D_2$). Treating regions 2 and 3 as an integrated demand zone, the import curve of this combined region ($I$) is given by the aggregate demand curves: $I=I_2+I_3=D_2+D_3=D$.

The market equilibrium for the integrated market is given by the solution to $E_1(p) = I(p) \iff S(p) = D(p).$ Figure \ref{fig:3_node_trade_fig} shows this as the intersection between the export/supply curve of zone 1 (solid red line) and aggregate import/demand curves in zones 2 and 3 (dashed blue line). The equilibrium price and quantity are given by $p^{**}$ and $q^{**}$, respectively. By symmetry, electricity imports in zone 3 and zone 2 from zone 1 both equal half of the market quantity. However, while zone 2 imports directly from zone 1, the imports in zone 3 from zone 1 necessarily flow through zone 2, given the configuration of the network (see Figure~\ref{fig:network_configurations}).

Suppose now that zone 2 restricts the interconnector capacity to zone 3 to zero. This effectively prevents producers in zone 1 to export electricity to consumers in zone 3, pushing zone 3 into autarky. Zones 1 and 2 can still trade with each other without restrictions. This leaves the export curve of zone 1 unaffected. The import curve of zone 2 also remains the same and, due to the capacity restriction to zone 3, this is the import curve faced by zone 1. In terms of Figure \ref{fig:3_node_trade_fig}, the import curve faced by zone 1 shifts from the dashed to the solid blue line. This yields a new equilibrium price and quantity for zones 1 and 2 of $\tilde{p}$ and $\tilde{q}$, respectively. Note in particular that $\tilde{p}<p^{**}$, i.e., the restriction decreases the electricity price in zones 1 and 2. For zone 3, the (extreme) assumption of no domestic supply implies that there does not exist a domestic market with a positive quantity and price.

The welfare effects of the restriction are as follows. Starting with zone 2, the lower price increases consumer surplus by the green area, which equals the increase in domestic welfare. In zone 1, the lower price decreases producer surplus (and thus, domestic welfare) by the green area plus the dotted blue area. In zone 3, the restriction decreases consumer surplus and domestic welfare by dotted orange area. The decrease in total system welfare (i.e., the deadweight loss) is represented by the dotted areas. Given the welfare increase in zone 2, the TSO in zone 2 has an incentive to limit the capacity on the border with zone 3. As this welfare increase is induced by a higher domestic price, we refer to this as the \textit{internal price mechanism}. Note, however, that the welfare increase in zone 2 comes at the expense of total welfare in the system.

While the current example illustrates that export restrictions in an importing region may increase domestic welfare through lowering the domestic price and thus, raising consumer surplus, it follows by an analogous argument that import restrictions in an exporting region may increase domestic welfare through raising the domestic price and thus, raising producer surplus. As the argument is completely symmetrical (if we replace all import nodes by export nodes and vice versa), we leave the details to the reader.

The examples in this section illustrate two mechanisms through which capacity restrictions by TSOs may increase domestic welfare: the price difference mechanism and the internal price mechanism. To facilitate the interpretation of our illustrations, we made use of relatively strong assumptions on the supply and demand (and thus import and export) curves. Nevertheless, the main principles underlying these mechanisms also apply in settings with more realistic assumptions. To analyse this further, the next section investigates the welfare effects of capacity restrictions in a setting that is representative for actual market conditions, with an electricity network consisting of many interconnected zones that are characterized by realistic supply and demand curves.

\section{Case study}
\label{sec:case_study}

In this section we study economic incentives for capacity reductions on interconnectors in a case study of the Northern-European day-ahead market for electricity. The main purpose of this case study is to investigate whether the two mechanisms we hypothesized in Section~\ref{sec:analytical_framework} can actually be observed in a realistic model and to quantify the welfare effects of the resulting capacity reductions.

\subsection{Model description}
\label{subsec:model_description}

\begin{figure}[t]
\centering
\includegraphics[width=\linewidth]{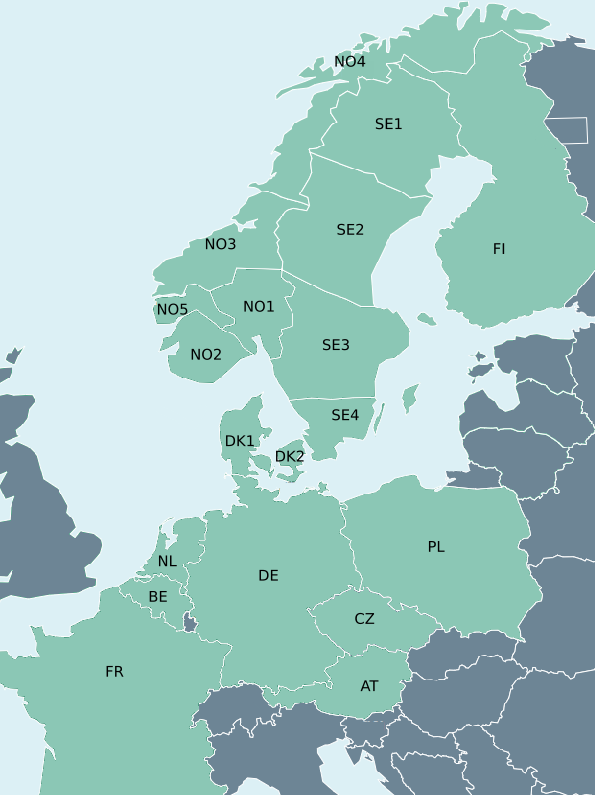}
\caption{Map of all price zones considered in the case study.}
\label{fig:map_price_zones}
\end{figure}

Before describing the model used in our case study we first describe the situation we aim to model. We consider part of the Northern-European transmission network. Specifically, we consider the twelve countries colored green in Figure~\ref{fig:map_price_zones} (Austria, Belgium, Czech Republic, Denmark, France, Finland, Germany, the Netherlands, Norway, Poland, and Sweden). These countries are integrated into the European electricity market, which is based on a zonal market structure. Each country is made up of one or several price zones. In each price zone electricity is traded in the day-ahead market at a single zonal price. Every time period, market participants log orders in the day-ahead market indicating how much they are willing to buy or sell at all potential prices. These orders are translated into an aggregate supply and demand curve per price zone. Matching these curves in all zones simultaneously, taking into account the possibility of transmitting electricity through interconnectors between different price zones, a market solution is found with a corresponding market clearing price in each price zone. Every actor in the market that indicated a willingness to buy or sell at this price is then bound to do so.

\begin{figure}[t]
\centering
\input{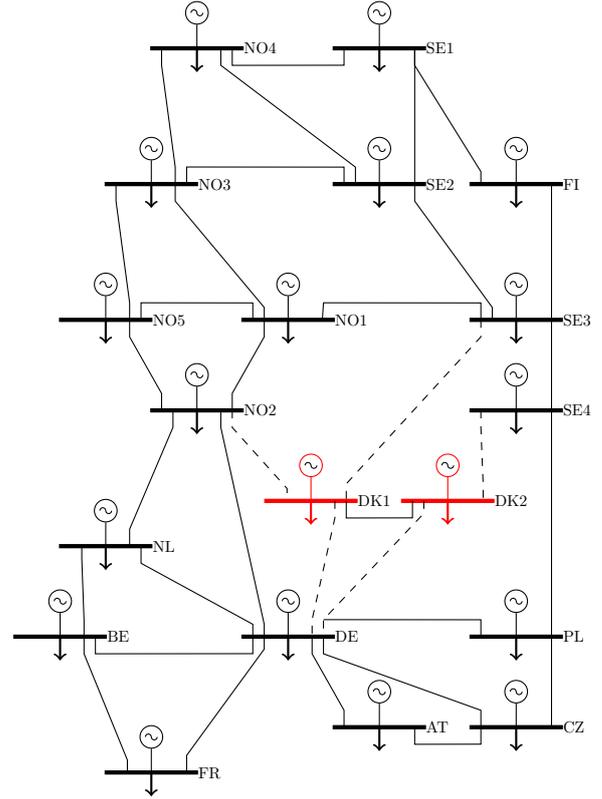}
\caption{Underlying graph representation of the power system used in the case study. Nodes representing Danish price zones are colored red; interconnectors from Danish zones to neighboring zones are indicated by dashed lines.}
\label{fig:network}
\end{figure}

Our model (which is based on the model in \cite{vanbeesten2022welfare}) is aimed to be a reflection of this day-ahead market. Here, we provide a high-level description of the model; a detailed mathematical formulation is provided in Section~\ref{sec:mathematical_formulation}. First, we define a graph, illustrated in Figure~\ref{fig:network}, that represents the underlying geographical network. Every node in the graph corresponds to a price zone in the day-ahead market, while every edge represents interconnector capacity between neighboring price zones. The actors in the day-ahead market are modeled as follows. Buyers on the market are represented by a representative consumer that tries to maximize its consumer surplus, defined as the area under its demand curve. The relevant demand curves are estimated based on historical consumption and price data. Sellers on the market are modeled as representative profit-maximizing generating companies without market power. The profit-maximization problem determines how much each generating company is willing to supply at each price, i.e., it implicitly determines a supply curve. Adding a market coupling constraint that matches net supply and net demand in each zone (and adding a ``dummy'' market operator that sets the flows on the interconnectors), these consumer and producer optimization problems together constitute an equilibrium model. Clearing the day-ahead market is equivalent to solving this equilibrium problem.

To solve the equilibrium problem, we first formulate it as a so-called mixed-complementarity model (MCP) \cite{gabriel2012} by combining the Karush-Kuhn-Tucker optimality conditions of all individual optimization models. It turns out that this MCP can be reformulated as a single quadratic optimization model, which can be solved using standard optimization packages such as Gurobi (see Section~\ref{sec:mathematical_formulation} in the appendix for more details). In our implementation, we solve for one week (i.e., 168 subsequent individual hours) simultaneously. When a solution is obtained, market-clearing prices are generated for every hour and corresponding welfare statistics can be computed straightforwardly. Here, we assume that the congestion rent earned on an interconnection between two price zones is equally distributed over the two TSOs responsible for those price zones; this is in line with typical agreements used in practice \cite{huang2016mind}.

To obtain a realistic representation of the Northern-European power market, we use historical data from this region as input data for our model. Specifically, we use historical data to (1) determine physical capacities of interconnectors, (2) determine production capacities of dispatchable generators, (3) obtain production patterns for solar and wind power, and (4) estimate demand curves based on estimated price elasticities and historical consumption and prices. For parameters that vary over time (e.g., renewable production), we use a sample of 100 historical weeks from the period 2016--2020. A full description of the data used in the case study is provided in Section~\ref{sec:data} in the appendix.


\subsection{Model validation} 
\label{subsec:model_validation}

From a structural point of view our model closely resembles the actual Northern-European electricity market. The zonal division and interconnectors match the actual market. Moreover, the market dynamics as modeled in our mathematical formulation reflect the design of the actual European day-ahead market: producers and consumers bid their supply and demand curves, respectively, after which a central market authority clears the market by picking prices that equate supply to demand in every zone. One structural limitation of our model is the fact that we only include a selection of European countries. However, as we focus on Denmark in our analysis, we made sure to include all of Denmark's neighboring countries and most of its neighbors' neighbors. Thus, we believe the resulting network to be a reasonable representation of the power market around Denmark.

Accurate parametrization of our model is a somewhat more challenging task. While demand curves can be estimated based on available consumption and price data and estimates for the price elasticity of demand (see Section~\ref{sec:data} in the appendix for more details), estimation of producers' supply curves (i.e., their marginal costs) is harder, since we are not aware of data on historical supply curves bid in the market. Instead, we estimate the marginal costs for each type of generator based on historical prices of coal and gas, carbon emission prices, and assumptions for electrical efficiencies (see Section~\ref{sec:data} in the appendix for more details). Though we made use of the most accurate data available, these estimates inevitably introduce some errors in the parametrization of the model.

\begin{figure*}
\begin{multicols}{2}
    \includegraphics[width=\linewidth]{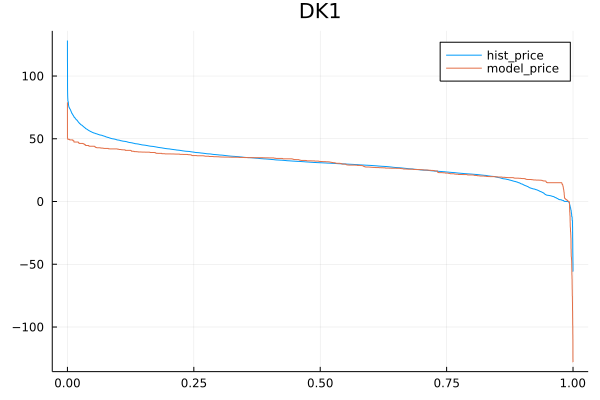}\par 
    \includegraphics[width=\linewidth]{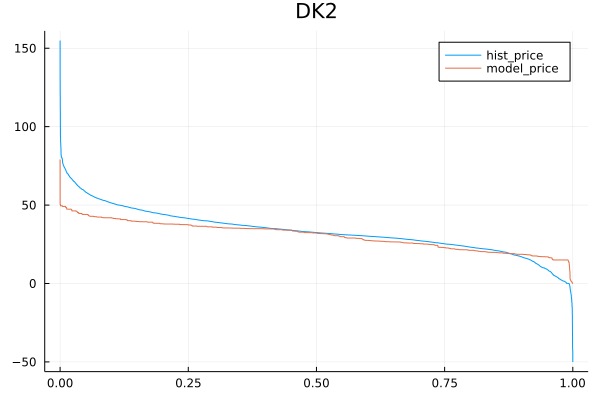}\par 
    \end{multicols}
\begin{multicols}{2}
    \includegraphics[width=\linewidth]{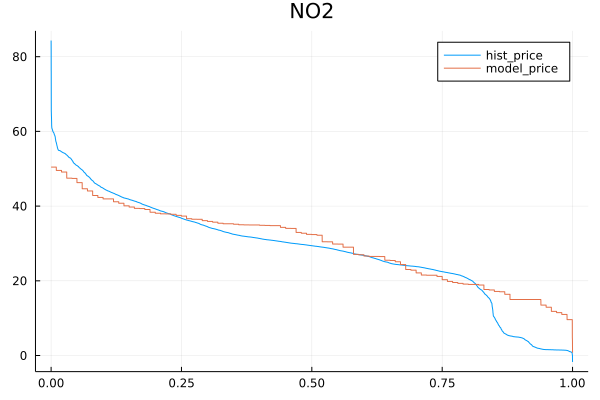}\par
    \includegraphics[width=\linewidth]{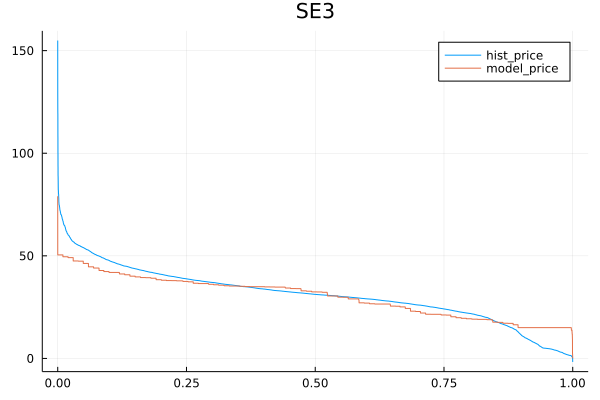}\par
\end{multicols}
\begin{multicols}{2}
    \includegraphics[width=\linewidth]{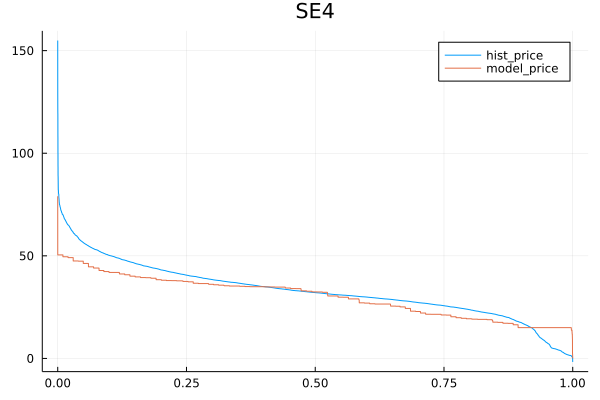}\par
    \includegraphics[width=\linewidth]{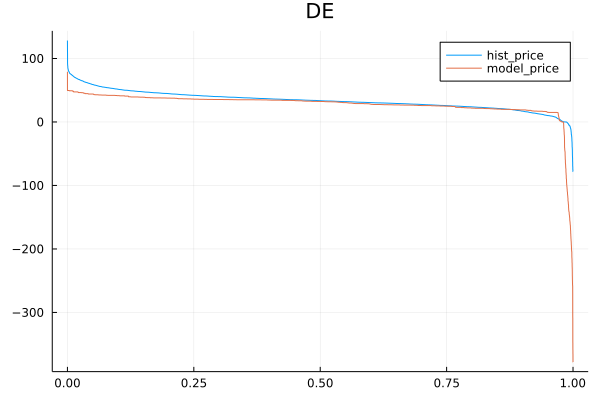}\par
\end{multicols}
\caption{Price-duration curves for both the model prices and historical prices in the most relevant price zones.}
\label{fig:price_duration_curves}
\end{figure*}

However, for our purposes we do not need an \textit{exact} representation of the historical power system. It suffices to populate our model with \textit{realistic} data, in the sense that they represent the overall dynamics in the power market to a sufficient degree. To test this, we plot so-called \textit{price-duration curves}. These curves show, for a given zone, the distribution of the price over all hours in all sampled weeks (sorted from high to low). In Figure~\ref{fig:price_duration_curves} we present the price-duration curves for the most relevant zones in our model: the Danish zones (DK1, DK2) and their neighboring zones. We observe that in all zones, the historical and the model price-duration curve are reasonably close. The most significant differences are observed in the tails of the distribution: in most zones, the tails of the distribution are somewhat more extreme in the historical data than in our model. A main reason for this may be the fact that in reality there is likely more variability in marginal costs than we capture in our data. As the welfare effects of restrictions tend to be larger when price differences are larger, this could imply that our results are somewhat underestimated. All in all, however, the price-duration curves show a good fit, suggesting that the model is a reasonable representation of the day-ahead market in the Northern-European power system. 

\subsection{Experimental design}
\label{subsec:experimental_design}

We give one country in the network the option of reducing the export capacities on its interconnections with foreign price zones. In this case study, we choose Denmark, consisting of price zones DK1 and DK2 (colored red in Figure~\ref{fig:network}). Denmark occupies a central location in the network we consider, functioning as a bottleneck for transmission between Scandinavia and mainland Europe. This makes it an interesting candidate to consider in our numerical study. 

We run a number of experiments to investigate whether the Danish TSO has incentives for capacity reduction and what the corresponding welfare effects are. Each experiment corresponds to a week of historical data. In total, we sample 100 weeks from the years 2016--2020. To obtain a balanced sample, we made sure to include 25 scenarios from every season (``spring'': Mar--May, ``summer'': Jun--Aug, ``autumn'': Sep--Nov, ``winter'': Dec--Feb). Each week is divided into hours, so every experiment consists of $7 \times 24 = 168$ time periods.

For every hour in the day-ahead market, the Danish TSO has the option of reducing the capacity on each of the interconnections between a Danish price zone and a neighboring price zone: DK1--NO1, DK1--NO2, DK1--SE3, DK1--DE, DK2--DE, DK2--SE4 (these correspond to the dashed lines in Figure~\ref{fig:network}). For each interconnection, we consider three possible capacity levels: 0\%, 50\%, and 100\% of the maximum capacity. Note that despite the European 70\% rule, capacity restrictions of this order of magnitude are not uncommon in Europe \cite{ACER2022report, efet2022insight}. We limit the number of capacity levels to only these three values for two reasons. First, in our model the Danish TSO essentially has perfect foresight: it knows the other market participants' actions beforehand and it can adapt its capacity decisions to it. This assumption introduces the risk of overstating the potential welfare gains that can be obtained through capacity reductions. By restricting the capacity decision to only three options, we take away much of the TSO's ability to exploit its perfect foresight. Essentially, it reduces to the ability to choose the best option out of the three capacity levels. We believe that it is reasonable to assume the TSO has this ability, as much of the day-ahead market conditions will be known to the TSO. Second, the limited number of capacity levels simplifies the process of solving our model. We simply perform a full enumeration of all possible capacity combinations and pick the combination with the highest associated Danish welfare. With five interconnections and three capacity levels, we have $5^3 = 243$ combinations, which is computationally feasible.

We run experiments in three different settings. First, the \textit{base case} corresponds to the setting described above: every hour the Danish TSO can choose between the capacity levels 0\%, 50\% and 100\% on every interconnection. Second, the \textit{long-term case} corresponds to a setting where the Danish TSO cannot choose different capacity levels in every hour, but must choose \textit{only one} set of capacity levels that holds for the foreseeable future. By comparing this case with the base case, we investigate how much additional welfare can be gained by adjusting the capacity reductions to the particular market circumstances in each week (thereby also providing a comparison between our paper and the literature on, e.g., transmission expansion planning, where one-time capacity expansion decisions must be made). For this purpose, we adapt our model slightly by including all weeks simultaneously and maximizing the \textit{expected} total Danish welfare, where every week is interpreted as a scenario with probability 1/100 = 1\%. Finally, the \textit{70\% case} corresponds to a setting with hourly capacity decisions (as in the base case) but capacity levels 70\%, 85\%, and 100\% for each interconnection. The purpose of this setting is to investigate how effective the European 70\% rule is at mitigating the strategic TSO behavior explored in this paper (assuming full compliance with the rule).

\section{Results}
\label{sec:results}

This section discusses the results of our numerical experiments. We discuss the base case, long-term case, and 70\% case, described above, one by one.

\subsection{Base case}
\label{subsec:base_case}

First, we consider the base case where the Danish TSO can choose different capacity levels for every hour, with possible capacity levels 0\%, 50\% and 100\% for each interconnector. Our focus in this subsection is on our main research question: does the Danish TSO have economic incentives to reduce transmission capacities on its international interconnectors? 

\begin{figure}[t]
    \centering
    \includegraphics[width=0.48\textwidth]{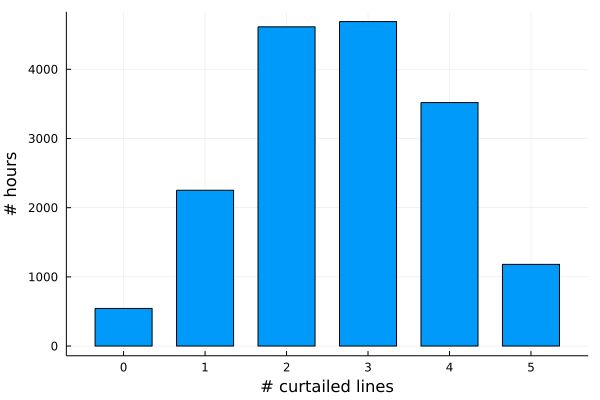}
    \caption{Histogram of the number of curtailed lines (i.e., interconnections with reduced capacity) in each hour in the base case.}
    \label{fig:hist_num_curt_lines}
\end{figure}

The short answer is yes: in the vast majority of cases it does. As can be observed from Figure~\ref{fig:hist_num_curt_lines}, in 96.8\% of all hours in the 100 weeks tested, the TSO is able to increase the Danish total welfare by means of capacity reductions. In most hours, two, three or four out of the five interconnections have their capacity reduced. Hence, in the overwhelming majority of cases, the TSO does have economic incentives for at least some level of transmission capacity reduction. In the remainder of this subsection we delve deeper into the magnitude of these capacity reductions, their resulting welfare effects, and their underlying mechanisms.

\begin{figure}[t]
    \centering
    \includegraphics[width=0.48\textwidth]{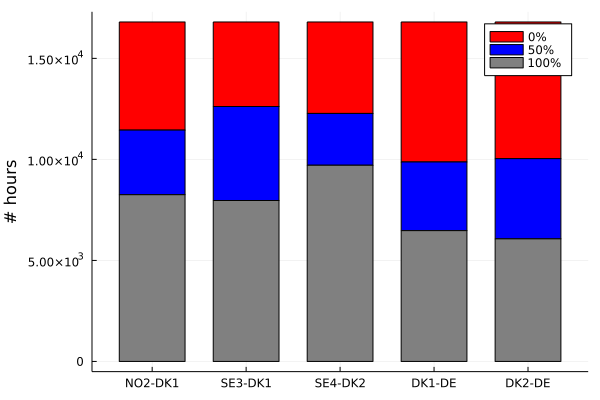}
    \caption{Plot of the level of available capacity of each line over all hours in the base case.}
    \label{fig:plot_curt_level}
\end{figure}

First, we investigate the \textit{magnitude} of capacity reductions. Figure~\ref{fig:plot_curt_level} illustrates the different levels of available capacity on every individual interconnection. We observe that all interconnections are used at full capacity in about 50\% of the hours.  In the other half of the hours, however, the interconnector capacity is reduced. Interestingly, we observe more hours where interconnector capacity is reduced all the way to zero than with more modest capacity reductions of 50\%. Hence, whenever incentives for capacity reductions exist, they tend towards extreme levels of reductions. This suggests that, if acted upon, these incentives have the potential to significantly disrupt the power market. Next,  Table~\ref{tab:avg_line_cap_usage} presents the average capacity availability of each interconnection. These range from as low as 48\% on the DK2--DE interconnection to 65\% for DK2--SE. We conclude that the observed levels of capacity reductions are quite significant.

\begin{table}[t]
    \centering
    \begin{tabular}{lc}
        \toprule
        Line & Avg. availability \\
        \midrule
        DK1--NO2 & 59\% \\
        DK1--SE3 & 61\% \\
        DK1--DE & 49\% \\
        DK2--SE4 & 65\% \\
        DK2--DE & 48\% \\
        \bottomrule
    \end{tabular}%
    \caption{Average capacity availability over all hours for each line connected to Denmark in the base case.}
    \label{tab:avg_line_cap_usage}
\end{table}

\begin{table}[]
    \centering
    \begin{tabular}{lllll}
        \toprule 
        Country & $\Delta$TW & $\Delta$CS & $\Delta$PS & $\Delta$CR \\
        \midrule 
        \textbf{DK} & \textbf{47} & \textbf{64} & \textbf{2} & \textbf{-19} \\
        NO & -54 & -273 & 219 & 0 \\
        SE & -57 & -269 & 219 & -8 \\
        FI & -22 & -33 & 25 & -14 \\
        DE & -21 & 41 & -47 & -15 \\
        NL & 3 & 1 & -1 & 3 \\
        BE & -1 & -4 & 3 & 0 \\
        FR & 0 & -10 & 11 & -2 \\
        AT & -5 & 6 & -11 & 0 \\
        CZ & -1 & -1 & 1 & -1 \\
        PL & 7 & -4 & 3 & 7 \\
        \midrule
        \textbf{Total} & \textbf{-105} & \textbf{-481} & \textbf{425} & \textbf{-49} \\
        \bottomrule
    \end{tabular}%
    \caption{Table of average welfare changes (expressed in millions of euros per year) for the base case.}
    \label{tab:welfare_effects}
\end{table}

Second, we study the \textit{welfare effects} of the optimal capacity reductions. Focusing on Denmark first, we see in Table~\ref{tab:welfare_effects} that the possibility of reducing transmission capacities leads to an average total welfare increase for Denmark of 47 million euros per year. To put this number in perspective, we compare it with the economic value of all power traded in Denmark\footnote{We prefer this approach over expressing the welfare change as a percentage of the total welfare itself, as the total welfare is ill-defined. To compute it, we need the area under the demand curve, for which we only have a domestic linear estimate based on historical prices and consumption levels. (Note that the welfare \textit{change} is well-defined, though, since the domestic estimate of the demand curve is sufficient to compute this.)}. We compute this value by multiplying all production and consumption in Denmark by the corresponding zonal prices, aggregating this over the entire time horizon, and dividing it by two (to correct for double counting of production and consumption). The Danish total welfare increase then amounts to $6\%$ of this economic value, which is relatively significant.



Other countries are also affected by the Danish capacity reductions. The countries that experience the strongest welfare effects are its neighbors Norway, Sweden, and Germany, as well as Finland. These countries are all negatively affected by the capacity reductions, mostly due to decreased consumer surplus (except for Germany, whose producer surplus is hurt). Most other, more distant countries are only moderately affected. Taken as a whole, the Northern-European power market suffers from the capacity reductions. The total welfare loss in all countries besides Denmark is 152 million euros per year, yielding a net welfare loss of 105 million per year for the system as a whole. This shows that the capacity reductions are indeed undesirable from a system perspective and can have significant detrimental effects.

\begin{figure}
    \centering
    \includegraphics[width=0.48\textwidth]{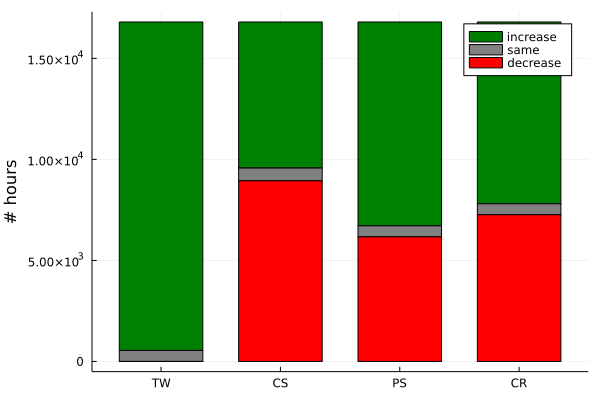}
    \caption{Plot indicating the direction of change for four Danish welfare measures over all hours in the base case.}
    \label{fig:plot_delta_dir}
\end{figure}

Finally, we move to the question of what \textit{mechanisms} cause the welfare increases in Denmark. In Section~\ref{sec:analytical_framework} we discussed two mechanisms: the price difference mechanism (related to an increase in congestion rent) and the domestic price mechanism (related to an increase in consumer surplus or producer surplus). To study which of these mechanisms is the main driving factor behind our results, we plot the direction of the change of Danish total welfare and its constituent parts in Figure~\ref{fig:plot_delta_dir}. While the total welfare increases in all hours with capacity reductions, the direction of change of its constituent parts does not show a clear pattern. Each constituent part increases in some hours, while it decreases in others. This suggests that the mechanism for welfare increase (through a higher domestic price, lower domestic price, or higher price differences with external zones) varies per hour. So none of the two mechanisms dominates the others, but the particular mechanism driving incentives for capacity reduction depends on the market circumstances at hand. 

\begin{figure*}
\centering
\begin{multicols}{2}
    \includegraphics[width=0.69\linewidth]{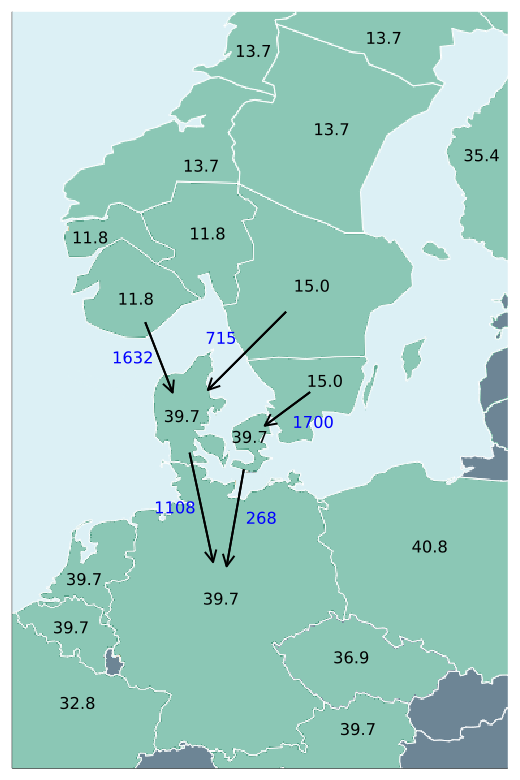}\par 
    \includegraphics[width=0.69\linewidth]{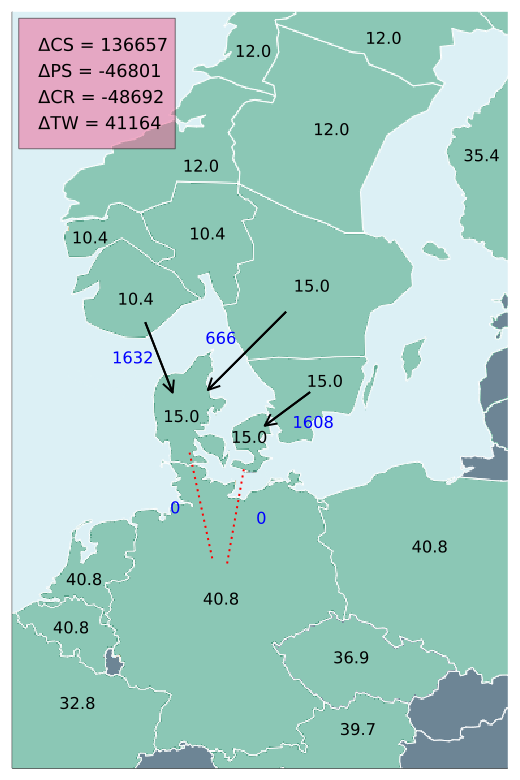}\par 
    \end{multicols}
\caption{Plot of the prices in every zone and the flow on every Danish interconnector for an example hour in the base case, without (left) and with (right) capacity reductions. Arrows indicate the direction of power flow. Solid black, dashed blue, and dotted red arrows correspond to 100\%, 50\% and 0\% available capacity, respectively. This plot shows period 114 in autumn week \#16, providing an example of a capacity reduction causing a lower domestic price, increasing consumer surplus.} 
\label{fig:map_CS}
\end{figure*}

\begin{figure*}
\centering
\begin{multicols}{2}
    \includegraphics[width=0.69\linewidth]{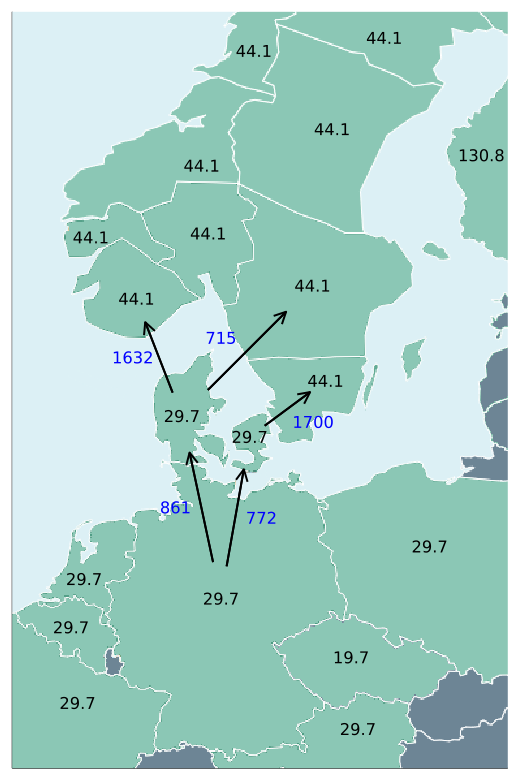}\par 
    \includegraphics[width=0.69\linewidth]{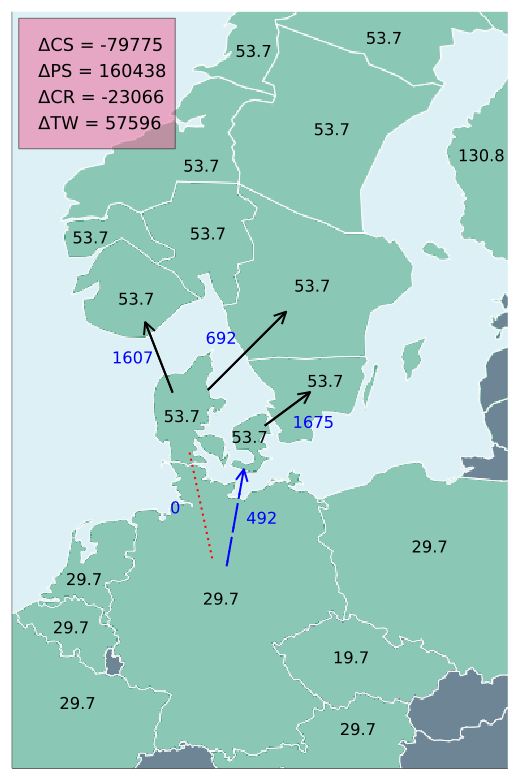}\par 
    \end{multicols}
\caption{Similar plot as in Figure~\ref{fig:map_CS}, but for period 77 in spring week \#10, providing an example of a capacity reduction causing a higher domestic price, increasing producer surplus.}
\label{fig:map_PS}
\end{figure*}

\begin{figure*}
\centering
\begin{multicols}{2}
    \includegraphics[width=0.69\linewidth]{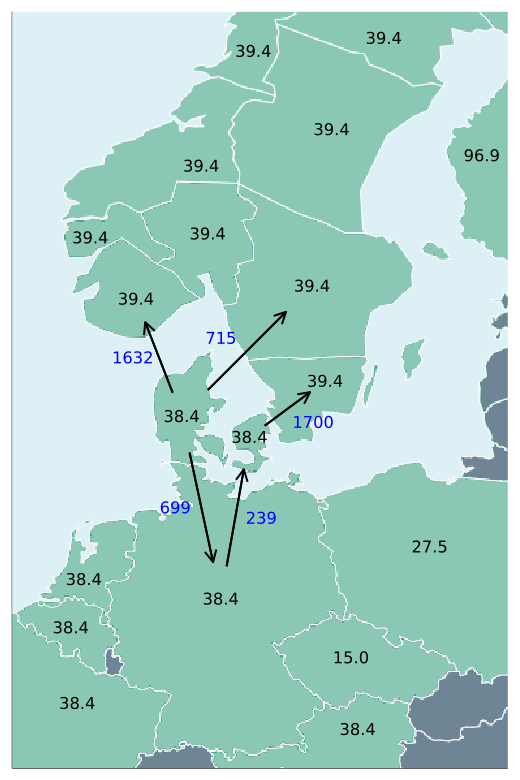}\par 
    \includegraphics[width=0.69\linewidth]{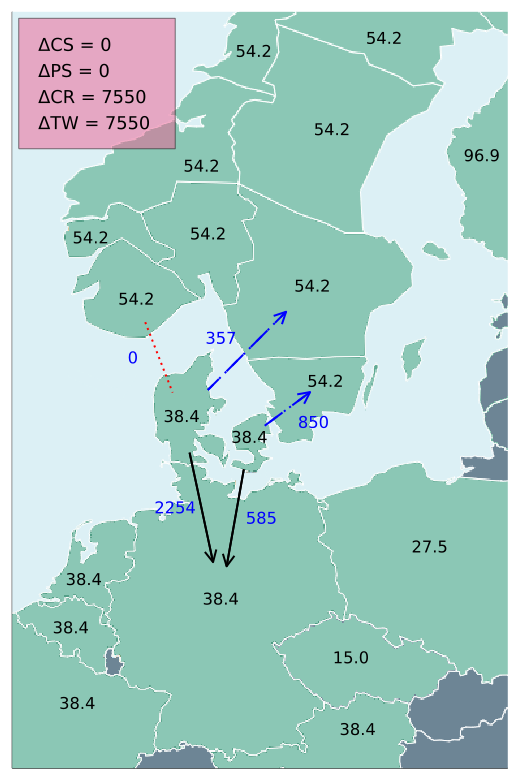}\par 
    \end{multicols}
\caption{Similar plot as in Figure~\ref{fig:map_CS}, but for period 119 in winter week \#14, providing an example of a capacity reduction causing higher price differences, increasing congestion rent.}
\label{fig:map_CR}
\end{figure*}

This finding is confirmed when we study individual hours. In Figure~\ref{fig:map_CS}--\ref{fig:map_CR} we present the results of three different individual hours that each achieve a welfare increase through different means. In Figure~\ref{fig:map_CS} a welfare increase is achieved by a lower domestic price, leading to a higher consumer surplus (while producer surplus and congestion rent are lower). In contrast, in Figure~\ref{fig:map_PS} a welfare increase is achieved through a higher domestic price, with the other two measures being lower, and in Figure~\ref{fig:map_CR} there is a larger price difference with external zones, leading to a higher congestion rent, but lower producer and consumer surplus. We conclude that indeed, different mechanisms for welfare increase may occur in different hours, depending on the particular market conditions at hand.

Nevertheless, the stark increase in Danish consumer surplus observed in  Table~\ref{tab:welfare_effects} indicates that the domestic price mechanism (particularly, through decreasing the price in Denmark) contributes most to the average welfare gains in Denmark. In practice, this means Denmark profits most from its ability to ``keep cheap electricity for its own consumers'' in periods where it would otherwise be exporting.

A final insight is found in terms of the effect of the capacity reductions on the flow of electricity to and from Denmark. For the hours in which the domestic price effect is dominant, the optimal capacity reductions tend to exaggerate Denmark's net import/export. That is, if Denmark imports more than it exports, it tends to limit export capacity and becomes an even stronger net importer and vice versa.  Specifically, this holds for 81\% of the hours in which congestion rent decreases as a result of capacity reductions (meaning that the price difference effect is absent and thus, the domestic price effect is responsible for the Danish welfare gain). This result is in line with the insight in Figure~\ref{fig:3_node_trade_fig}, where the importing node 2 profits from a decreasing price and the resulting increase in net import.

\subsection{Long-term case}
\label{subsec:long-term_case}

Next, we consider the \textit{long-term case}, where the Danish TSO can choose only one set of capacity levels for the entire foreseeable future, with possible capacity levels 0\%, 50\%, and 100\% for each interconnector. Our focus in this subsection is on the question how much the additional flexibility in the base case (in terms of adjusting the capacity levels to the particular weekly market circumstances) contributes to economic incentives for capacity reductions and to the corresponding welfare effects. This also provides a comparison of the results in our paper with the literature on, e.g., transmission expansion planning, where long-term capacity level decisions are made.

The results show that in terms of maximizing Danish total welfare it is optimal to reduce the capacity of the interconnection between DK1 and SE3 to 50\%, while keeping all other capacities at 100\%. Hence, from a long-term perspective limiting capacity on the SE3--DK1 is desirable for Denmark. Put differently, if the interconnection didn't exist and a transmission expansion plan were to be developed, it would be in Denmark's interest to only invest in an interconnection with limited capacity. In our setting we do not consider increasing transmission capacity, however, but \textit{decreasing} it, i.e., this long-term setting may be interpreted as a reverse transmission expansion planning problem.

Compared with the base case, we make one very important observation. Even though there is no incentive to structurally limit interconnection capacity on most interconnections (as shown by the long-term case results), if we allow the capacity reductions to depend on the (hourly) market conditions, we observe incentives for capacity reductions on \textit{all} interconnections. This result highlights the importance of considering the dynamic aspect of this problem.

The expected welfare effects of the optimal 50\% capacity reduction on the SE3--DK1 interconnection are shown in Table~\ref{tab:welfare_effects:long-term}. The expected total welfare increase for Denmark is 4 million euros per year. This is only a fraction of the 47 million observed in the base case. This finding provides strong evidence that the additional flexibility in the base case greatly increases the ability of the Danish TSO to increase Denmark's welfare through capacity reductions. Again, it emphasizes the importance of considering this flexibility. The expected welfare decrease for the system as a whole is 36 million euros per year. This is also a reduction compared with the 105 million decrease in the base case, but a much smaller reduction than on the effect on Denmark itself. 

Table~\ref{tab:welfare_effects:long-term} also provides insight into the welfare-increasing mechanism at play in the capacity reduction on the SE3--DK1 line. The capacity reduction increases consumer surplus, while decreasing producer surplus. This suggests that it yields a lower price on average, which in turn suggests that the capacity reduction tends to limit \textit{exports} to Sweden. The increase in consumer surplus does not compensate for the decrease in producer surplus, though. The deciding factor is the increase in congestion rent. Without this effect, the capacity reduction would not have been desirable. This provides evidence that the mechanism at work is the price difference mechanism (cf. Section~\ref{subsec:2node}).

\begin{table}[]
    \centering
    \begin{tabular}{lllll}
        \toprule 
        Country & $\Delta$TW & $\Delta$CS & $\Delta$PS & $\Delta$CR \\
        \midrule 
        \textbf{DK} & \textbf{4} & \textbf{32} & \textbf{-41} & \textbf{13} \\
        NO & 22 & -184 & 167 & 38 \\
        SE & -30 & -179 & 156 & -8 \\
        FI & -12 & -1 & 1 & -12 \\
        DE & -23 & 62 & -63 & -22 \\
        NL & 2 & -2 & 1 & 2 \\
        BE & 0 & -1 & 1 & 0 \\
        FR & -2 & 6 & -7 & -2 \\
        AT & 1 & 7 & -6 & -1 \\
        CZ & -1 & 0 & 0 & -1 \\
        PL & 3 & 0 & 0 & 3 \\
        \midrule
        \textbf{Total} & \textbf{-36} & \textbf{-259} & \textbf{211} & \textbf{12} \\
        \bottomrule
    \end{tabular}%
    \caption{Table of average welfare changes (expressed in millions of euros per year) for the long-term case.}
    \label{tab:welfare_effects:long-term}
\end{table}

\subsection{70\% case}
\label{subsec:70_percent_case}

Finally, we consider the \textit{70\% case}, where the Danish TSO can choose different capacity levels for every hour (as in the base case), but the possible capacity levels for each interconnector are 70\%, 85\%, and 100\%, reflecting full compliance with the European 70\% rule. Our focus in this subsection is on the question whether the 70\% rule (if fully complied with) is effective at mitigating the incentives for capacity reduction and the resulting consequences observed in the base case.

\begin{figure}
    \centering
    \includegraphics[width=0.48\textwidth]{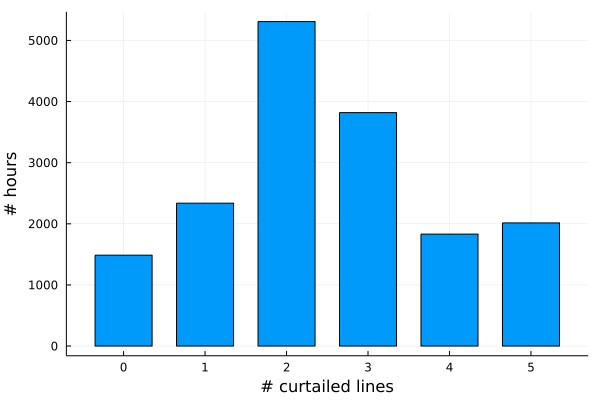}
    \caption{Histogram of the number of curtailed lines (i.e., interconnections with reduced capacity) in each hour in the 70\% case.}
    \label{fig:hist_num_curt_lines:limited}
\end{figure}

In Figure~\ref{fig:hist_num_curt_lines:limited} we observe that in 91.1\% of the hours in the 100 weekly samples the Danish TSO limits transmission capacity on at least one interconnector. This means that, compared with the base case, the amount of hours without capacity reductions has more than doubled (from 3.2\% to 8.9\%). Hence, the inability to impose drastic capacity reductions removes the incentive to reduce transmission capacity in some situations. Nevertheless, incentives for capacity reductions continue to exist in the vast majority of hours. Thus, we conclude that the 70\% rule is not effective at removing the incentives for capacity reductions altogether.

\begin{figure}
    \centering
    \includegraphics[width=0.48\textwidth]{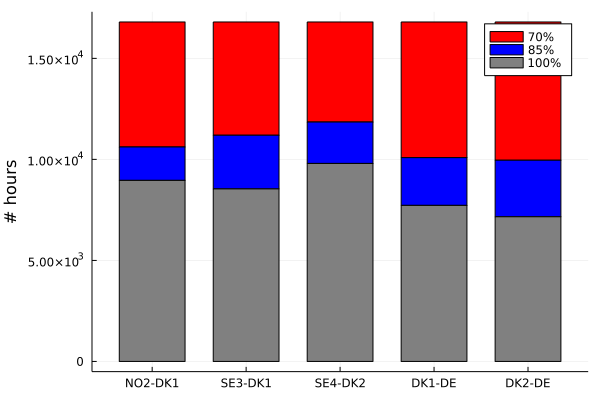}
    \caption{Plot of the level of available capacity of each line over all hours for the 70\% case.}
    \label{fig:plot_curt_level:limited}
\end{figure}

Figure~\ref{fig:plot_curt_level:limited} shows that, as in the base case, capacity reductions on the different interconnectors show very similar patters. Each interconnector has its capacity reduced in approximately half of the hours. Moreover, the more extreme reductions (to 70\%) are much more common than the milder restrictions (to 85\%). This provides an indication that in many cases the 70\% rule is binding, i.e., that the TSO would wish to reduce capacities further if it could. The restrictions result in the average capacity availabilities presented in Table~\ref{tab:avg_line_cap_usage:limited}. Clearly, the average availabilities are much higher than in the base case; the values increase from 48--65\% to 85--89\%.  

\begin{table}[]
    \centering
    \begin{tabular}{lc}
        \toprule
        Line & Avg. availability \\
        \midrule
        DK1--NO2 & 87\% \\
        DK1--SE3 & 88\% \\
        DK1--DE & 86\% \\
        DK2--SE4 & 89\% \\
        DK2--DE & 85\% \\
        \bottomrule
    \end{tabular}%
    \caption{Average capacity availabiliy over all hours for each line connected to Denmark in the 70\% case.}
    \label{tab:avg_line_cap_usage:limited}
\end{table}

In Table~\ref{tab:welfare_effects:limited_case} we present the welfare effects resulting from capacity reductions in the 70\% case. We observe an average gain in Danish total welfare of 13 million euros per year. This welfare gain is 71\% smaller than in the base case. In contrast with the base case, increases in congestion rent now constitute a significant part of the welfare gains, pointing at an increased importance of the price difference mechanism. For the system as a whole, the capacity reductions yield an average total welfare loss of 35 million euros per year, a reduction of 67\% compared to the base case. From these results we can infer that the 70\% rule, if fully complied with, substantially frustrates Denmark's ability to gain welfare by capacity reductions and mitigates the resulting adverse effects on other countries. This can be interpreted as an argument for enforcement of the 70\% rule, which is especially relevant given the fact that violations of this rule are commonplace \cite{ACER2022report, efet2022insight}.

Nevertheless, even within this setting of full compliance with the 70\% rule, the capacity reductions lead to a yearly welfare loss for all other countries combined of 49 million euros per year. Zooming in on particular groups, the effects are even more significant. For instance, Norwegian consumers lose 84 million euros in welfare per year due to increased electricity prices. Hence, even though the effects are dampened by the 70\% rule, significant welfare effects continue to exist. 

\begin{table}[t]
    \centering
    \begin{tabular}{lllll}
        \toprule 
        Country & $\Delta$TW & $\Delta$CS & $\Delta$PS & $\Delta$CR \\
        \midrule 
        \textbf{DK} & \textbf{13} & \textbf{15} & \textbf{-10} & \textbf{8} \\
        NO & -19 & -84 & 63 & 3 \\
        SE & -21 & -77 & 59 & -3 \\
        FI & -6 & 3 & -3 & -6 \\
        DE & -3 & 21 & -25 & 2 \\
        NL & 0 & 0 & 0 & 0 \\
        BE & 0 & -2 & 1 & 0 \\
        FR & 0 & -3 & 4 & -1 \\
        AT & -2 & 2 & -4 & 0 \\
        CZ & 0 & 0 & 0 & 0 \\
        PL & 2 & -1 & 1 & 2 \\
        \midrule
        \textbf{Total} & \textbf{-35} & \textbf{-127} & \textbf{87} & \textbf{5} \\
        \bottomrule
    \end{tabular}%
    \caption{Table of average welfare changes (expressed in millions of euros per year) for the 70\% case.}
    \label{tab:welfare_effects:limited_case}
\end{table}

\section{Conclusion}
\label{sec:conclusion}

We consider a zonal international power market and investigate potential economic incentives for TSOs to reduce transmission capacities on interconnectors in the day-ahead market. In contrast with the (limited) literature on this topic, which focuses on the possibility of avoiding expected balancing costs by means of capacity reductions in the day-ahead market, we ignore balancing costs and focus exclusively on immediate welfare effects in the day-ahead market itself.

First, we use an analytical framework to identify two different mechanisms by which domestic welfare can be increased through capacity reductions on interconnectors. The first mechanism, called the price difference mechanism, is based on manipulating price differences with neighboring price zones. This mechanism results from the practice of equally splitting congestion rent between neighboring TSOs and can already be observed in a two-zone setting, which contrast with conventional insights from the trade literature. The second mechanism, called the domestic price mechanism, is based on manipulating the domestic price in order to increase either consumer surplus or producer surplus. This mechanism is only observed in networks with at least three zones. 

Second, we run numerical experiments on a case study with realistic data from the Northern-European day-ahead market to investigate whether we can actually observe these economic incentives. Taking the Danish TSO as an example, we find that in 97\% of the historical hours tested, the TSO indeed has an incentive to reduce the transmission capacity on its interconnectors. On average, most lines are limited to about 60\% of their capacity on average, with the lines to Germany limited to less than half their capacity. This leads to an average welfare gain of 47 million euros per year for Denmark, while causing a yearly welfare loss of 105 million euros for the system as a whole. The prevailing mechanism that is responsible for the incentives for capacity reduction turns out to depend heavily on the specific market situation at hand. All in all, these results suggest that incentives for capacity reductions in the day-ahead market may well exist in practice and, if acted upon, can have significant welfare effects.

To contrast our results with the literature on long-term decisions for interconnector capacities (e.g., the transmission expansion planning and market coupling literature), we run an experiment where only a single decision for capacity reduction can be made that holds for all hours simultaneously. In this setting, only one interconnection (DK1--SE3) is restricted to 50\% of its capacity. The resulting welfare gain in Denmark is only 8\% of the welfare gain in the base case with hourly capacity restrictions. Moreover, the adverse welfare effect on the system as a whole reduces by two thirds. These results highlight that the issue studied in this paper differs significantly from the long-term decision problems typically studied in the literature.

To explore how regulation may mitigate the capacity reduction behavior and its welfare effects studied in this paper, we ran an additional set of experiments in which we enforce the European 70\% rule, stating that the available capacity on each interconnection should always be at least 70\%. Under this rule, the welfare effects (both on Denmark and on the system as a whole) are indeed reduced by approximately two thirds. Nevertheless, capacity restrictions still occur in 91\% of all historical hours tested and the welfare effects, though significantly reduced, remain substantial. We conclude that the 70\% rule can help mitigate the adverse effects of capacity reductions (thus, providing an argument for enforcing the rule), but is not sufficient to eliminate the problem completely.

Future research may be aimed at tackling some of the limitations of our paper. One limitation is that our analysis does not capture \textit{interactions} between different TSOs. We only allow a single TSO to act in a welfare-maximizing way, while keeping all other TSOs passive. In reality, the interaction between different TSOs may cause behavior not captured in our model. For one, every hour constitutes a simultaneous game where not just one, but also all other TSOs can restrict interconnection capacity to increase their domestic welfare. This might change the optimal capacity reduction decisions for each individual TSO. Moreover, the subsequent hours are not independent, but constitute a repeated game, in which TSOs might retaliate if another TSO acts too ``selfishly''. This limits the extent to which our results translate directly into behavior we expect to observe in practice. 
To address this limitation, it would be useful to run numerical experiments in a setting where multiple TSOs have the ability to simultaneously limit transmission capacity on interconnectors. This requires more advanced modeling approaches that can capture the interactions between different TSOs (e.g., equilibrium problems with equilibrium constraints (EPEC) \cite{gabriel2012}). Moreover, it would be interesting to use game-theoretical approaches to gain more insight into capacity restrictions from a long-term perspective, focusing in particular on possible retaliatory behavior.

A second limitation is that, even though we make use of detailed historical data, our model remains an approximation and the results are therefore also approximations rather than empirical estimates. Furthermore, our analysis implies that economic incentives for capacity reductions exist, but we have not analyzed the extent to which such restrictions are implemented in practice.
To address this issue, it would be interesting to \textit{empirically} investigate whether incentives not only exist in theory, but are also acted upon in practice. This is a challenging empirical question, since the motivations for TSO behavior are not directly observable. Hence, careful attention should be paid to the methodologies used to answer this question. Such investigations would contribute to a better understanding of the obstacles to efficient utilization of interconnector capacities. 


Finally, an interesting direction for future research would be to study the interaction between the economic incentives for capacity recuction on interconnectors studied in this paper and those studied in the literature. In particular, we focus on welfare gains in the day-ahead market, while the literature focuses mainly on the avoidance of balancing cost in the balancing market. It would be interesting to learn whether these benefits tend to go hand in hand or whether there is a tradeoff between the two.

\paragraph{CRediT authorship contribution statement}
\textbf{E. Ruben van Beesten:} Conceptualization, Methodology, Formal Analysis, Writing -- Original Draft. \textbf{D. Hulshof:} Conceptualization, Methodology, Formal Analysis, Data Curation, Writing -- Original draft.

\paragraph{Funding}
Ruben van Beesten's work was supported by FME MoZEES.

\paragraph{Declaration of Competing Interest}
The authors declare that they have no known competing financial interests or personal relationships that could have appeared to influence the work reported in this paper.

\paragraph{Acknowledgements}
We wish to thank the participants of the 30th Young Energy Economists and Engineers Seminar in Copenhagen for their valuable input; in particular Ulrike Pfefferer and Karyn Morrissey, who reviewed an early version of this paper. Moreover, we thank Raquel Alonso Pedrero, Naser Hashemipour, Pedro Crespo del Granado, and Asgeir Tomasgard for their comments on a later draft.

\begin{appendices}

\renewcommand{\theequation}{A.\arabic{equation}}

\section{Mathematical formulation}
\label{sec:mathematical_formulation}
In this section we describe the mathematical model used for our numerical experiments in Section~\ref{sec:case_study}. The model and its description in this section are based in part on the model from \cite{vanbeesten2022welfare}.

\subsection{Notation}

\noindent\textbf{Sets:}

\renewcommand{\arraystretch}{1.5}
\begin{tabularx}{0.95\linewidth}{@{}>{\bfseries}l@{\hspace{.5em}}X@{}}
    $\mathcal{N}$ & Set of price zones (i.e., nodes in the network), indexed by $n$ \\
    $\mathcal{L}$ & Set of lines (i.e., interconnections), indexed by $l$ \\
    $\mathcal{G}$ & Set of dispatchable generator types, indexed by $g$ \\
    $\mathcal{T}$ & Set of time periods (i.e., hours), indexed by $t$ \\
\end{tabularx}
\pagebreak

\noindent\textbf{Parameters:}

\renewcommand{\arraystretch}{1.5}
\begin{tabularx}{0.95\linewidth}{@{}>{\bfseries}l@{\hspace{.5em}}X@{}}
    $R_{nt}$ & Production from renewables in zone $n$ in period $t$ [$\SI{}{\mega\watt h}$] \\
    $C_{gt}$ & Marginal cost for dispatchable generation of type $g$ in period $t$ [$\SI{}{\text{\euro}\per\mega\watt h}$]\\
    $G_{ng}$ & Generation capacity for generator type $g$ in zone $n$ [$\SI{}{\mega\watt}$]\\
    $Q_{ng}$ & Production limit for generation of type $g$ in zone $n$ over the planning horizon [$\SI{}{\mega\watt h}$]\\
    $A_{nl}$ & Zone-line incidence matrix entry for zone $n$ and line $l$\\
    $F_{lt}$ & Capacity of line $l$ in time period $t$ [$\SI{}{\mega\watt}$] (for the lines connected to Denmark, this parameter is varied by the Danish TSO in the experiments)\\
    $D^{A}_{nt}$ & Slope of inverse demand function for zone $n$ in time period $t$\\
    $D^{B}_{nt}$ & Intercept of inverse demand function for zone $n$ in time period $t$\\
\end{tabularx}
\\

\noindent\textbf{Variables:}

\begin{tabularx}{0.95\linewidth}{@{}>{\bfseries}l@{\hspace{.5em}}X@{}}
    $q_{ngt}$ & Production by generator type $g$ in zone $n$ in time period $t$ [$\SI{}{\mega\watt h}$] \\
    $f_{lt}$ & Flow in line $l$ in time period $t$ [$\SI{}{\mega\watt h}$] \\
    $d_{nt}$ & Demand (i.e., consumption) in zone $n$ in time period $t$ [$\SI{}{\mega\watt h}$] \\
    $\pi_{nt}$ & Price in zone $n$ in time period $t$ [$\SI{}{\text{\euro}\per\mega\watt h}$] 
\end{tabularx}

\subsection{MCP formulation}
\label{subsec:MCP_formulation}

As explained in Section~\ref{sec:case_study}, our model is an equilibrium model in which all market participants act simultaneously. We take a time horizon of one week with an hourly time resolution. Every period (i.e., hour), each market participant solves its individual optimization problem (where we assume that all actors are price takers). Moreover, the actors are linked together using a market clearing constraint. 

Mathematically, this constitutes a  mixed-complementarity model (MCP), consisting of the Karush-Kuhn-Tucker (KKT) conditions of the individual optimization problems of all market participants plus a market clearing constraint. Rather than presenting the MCP itself, we present the individual optimization problems whose KKT conditions define the MCP.

\subsubsection{Dispatchable power producer problem}
\label{subsubsec:producer_problem}

In every zone $n \in \mathcal{N}$, all dispatchable power production resources are aggregated to a single profit-maximizing producer. The producer can freely choose its generation levels to maximize its profits. Eq.~\eqref{opt:GenCoProblem} describes the optimization problem for the generator located in zone $n \in \mathcal{N}$, where the constraints are defined $\forall \: g \in \mathcal{G}, \  t \in \mathcal{T}$.

\begin{maxi!}[1]
    {q_{ngt}}
    {\sum_{g \in \mathcal{G}}\sum_{t \in \mathcal{T}} \left( \pi_{nt} - C_{gt} \right) q_{ngt} \label{opt:GenCoObjective}}
    {\label{opt:GenCoProblem}}
    {}
    \addConstraint{q_{ngt}}{\leq G_{ng}\label{opt:GenCoProdCap}}
    \addConstraint{\sum_{t\in\mathcal{T}} q_{ngt}}{\leq Q_{ng}\label{opt:GenCoEnergyLimit}}
    \addConstraint{q_{ngt}}{\geq 0\label{opt:GenCoLargerThanZero}}
\end{maxi!}
The objective function in Eq.~\eqref{opt:GenCoObjective} consists of maximizing the expected revenue minus production cost, which are both assumed to be linear in the production level $q_{n g t}$. Eq.~\eqref{opt:GenCoProdCap} states that the production of each generation type must not exceed the corresponding capacity. Eq.~\eqref{opt:GenCoEnergyLimit} states that the total production over the planning horizon must be no more than the available quantity. The purpose of this constraint is to model the amount of available water for hydropower production. Finally, Eq.~\eqref{opt:GenCoLargerThanZero} states that the generation quantities must be non-negative.

\subsubsection{Consumer problem}
\label{subsubsec:consumer_problem}

In every zone $n \in \mathcal{N}$, the preferences of the consumers are represented by a linear demand curve. The fact that the consumers' behavior follows the demand curve can equivalently be stated by saying that the consumers maximize the consumer surplus, i.e., the area under the demand curve \cite{vanbeesten2022welfare}. Eq.~\eqref{opt:ConsumerProblem} represents the corresponding maximization problem for the consumers located in zone $n$.

\begin{maxi!}[1]
    {d_{ nt}}
    {\sum_{t\in\mathcal{T}}\left(\frac{1}{2} D^A_{nt} d_{nt} + D^B_{nt} - \pi_{nt} \right) d_{nt}\label{opt:ConsumerObjective}}
    {\label{opt:ConsumerProblem}}
    {}
\end{maxi!}
Eq.~\eqref{opt:ConsumerObjective} is the objective function for the consumers, representing the consumer surplus.

\subsubsection{Market operator problem}
\label{subsubsec:TSO_problem}
As a third actor in our model, we introduce a market operator, whose task is to set the flows $f_{lt}$ on the transmission lines. 
The market operator should act as a dummy player that simply clears the market based on the supply and demand curves implied by the producer and consumer problems. To this end, it sets the flows in such a way that the market is cleared in a social welfare-maximizing way, while satisfying transmission capacity constraints on interconnectors.\footnote{This is in line with the guiding principles of the Euphemia algorithm used in the coupled European day-ahead markets, which \textit{``has to decide which orders are to be executed and which orders are to be rejected in concordance with the prices to be published such that: (1) the social welfare (consumer surplus + producer surplus + congestion rent across the regions) generated by the executed orders is maximal; (2) the power flows induced by the executed orders, resulting in the net positions do not exceed the capacity of the relevant network elements''} \cite{euphemia}.}
This can be modeled by assuming the market operator is a price taker and that it follows the following two guidelines: (1) electricity flows from low price zones to high price zones and (2) as long as a price difference between two nodes persists, flow on the line between those nodes should be increased as much as possible (cf. \cite{vanbeesten2022welfare}). In other words, the market operator should maximize the product of the price difference and the flow on each line (ignoring the effect of the flows on prices, cf. the price-taker assumption), i.e., it should maximize congestion rent. Hence, the optimization problem for the market operator is given by 
%
%
\begin{maxi!}[1]
    {f_{lt}}
    {-\sum_{n \in \mathcal{N}}\sum_{l\in \mathcal{L}}\sum_{t \in \mathcal{T}} A_{nl} f_{lt} \pi_{nt}\label{opt:TSOObjective}}
    {\label{opt:TSOProblem}}
    {}
    \addConstraint{f_{lt}}{\leq F_{lt}\label{opt:TSOflowPos}}
    \addConstraint{f_{lt}}{\geq -F_{lt}\label{opt:TSOflowNeg}}
\end{maxi!}
The objective function in Eq.~\eqref{opt:TSOObjective} consists of the congestion rent earned from all lines. Eqs.~\eqref{opt:TSOflowPos} and \eqref{opt:TSOflowNeg} state that the flow in a line must not exceed its capacity. Note that in our experiments, we vary the values of the capacities $F_{lt}$ for lines $l \in \mathcal{L}$ connected to Denmark. Thus, the Danish TSO's ability to affect interconnector capacity is reflected in the fact that it can affect some of the parameters $F_{lt}$.

\subsubsection{Market clearing}
\label{subsubsec:market_clearing}
Finally, we add a market clearing constraint that is used to connect all the market actors' decisions together. It guarantees that in every node and every time period, the market clears, i.e., that net supply meets net demand. 
For every $n \in \mathcal{N}$ and $t \in \mathcal{T}$, it is given by
\begin{equation}
    d_{nt} + \sum_{l\in \mathcal{L}} A_{nl} f_{lt} = \sum_{g\in \mathcal{G}} q_{ngt} + R_{nt}  \label{opt:MarketClearing}  
\end{equation}
In particular, Eq.~\eqref{opt:MarketClearing} states that the sum of demand and net outgoing flows must be equal to the total amount of power generated from both conventional and renewable sources. The market price $\pi_{nt}$ is the dual variable corresponding to this constraint.

\subsection{Quadratic programming reformulation}
\label{subsec:quadratic_programming_reformulation}

In line with the classical result by \cite{samuelson1952}, it can be shown that the MCP formed by the KKT conditions corresponding to \eqref{opt:GenCoProblem}--\eqref{opt:TSOProblem} in conjunction with the market clearing constraint \eqref{opt:MarketClearing} is equivalent to a central planner quadratic optimization problem in which total welfare is maximized. The proof of this equivalence, which we omit for brevity, is through the observation that the KKT conditions to the quadratic program, which are necessary and sufficient, are equivalent to the MCP defined above. The quadratic program is given by  Eq.~\eqref{opt:CentralPlanner}, where the constraints are defined $\forall \: n \in \mathcal{N}, \ g \in \mathcal{G}, \ l \in \mathcal{L}, \ t \in \mathcal{T}$.

\begin{maxi!}[1]
    {\substack{q_{ngt}, \  d_{nt}, \  f_{lt}}}
    {\sum_{n \in \mathcal{N}}\sum_{t \in \mathcal{T}}\left(\frac{1}{2} D^A_{nt} d_{nt} + D^B_{nt} \right) d_{nt}}{\label{opt:CentralPlanner}}{} \nonumber
    \breakObjective{-\sum_{n\in\mathcal{N}} \sum_{g\in\mathcal{G}}\sum_{t\in\mathcal{T}} C_{gt} q_{ngt}\label{opt:CentralPlannerObj3}} 
    \addConstraint{q_{ngt}}{\leq G_{ng} \label{opt:CentralPlannerConFirst}}
    \addConstraint{\sum_{t\in\mathcal{T}} q_{ngt}}{\leq Q_{ng}}
    \addConstraint{d_{nt} + \sum_{l\in \mathcal{L}} A_{nl} f_{lt}}{=\sum_{g\in \mathcal{G}}q_{ngt} + R_{nt}}
    \addConstraint{f_{lt}}{\leq F_{lt}}
    \addConstraint{f_{lt}}{\geq -F_{lt}}
    \addConstraint{q_{ngt}, d_{nt}}{\geq 0}\label{opt:CentralPlannerConLast}
\end{maxi!}
Here, the objective function in Eq.~\eqref{opt:CentralPlannerObj3} is equal to the sum of the objective functions of all market participants' optimization problems (see \cite{vanbeesten2022welfare}). The constraints in Eqs.~\eqref{opt:CentralPlannerConFirst}--\eqref{opt:CentralPlannerConLast} are a concatenation of the constraints from all market actors' individual optimization problems and the market clearing constraint. 

The optimization problem in Eq.~\eqref{opt:CentralPlanner} can be solved using off-the-shelf quadratic programming solvers. We implement the model in Julia using JuMP \cite{DunningHuchetteLubin2017} and solve the model using the commercial solver Gurobi \cite{gurobi_new}.

\section{Data}
\label{sec:data}

In this section we describe the data used to parametrize the model from Appendix~\ref{sec:mathematical_formulation} used in the numerical experiments in Section~\ref{sec:case_study}. The data set spans the period 2016--2020 and contains a mixture of hourly, daily, monthly and annual series.

The underlying network (illustrated in Figure~\ref{fig:network}) is based on the actual structure of the electricity market in Northern-Europe, with a limited number of connected price zones. The aggregate physical capacities of the interconnectors between price zones come from the following sources: the EMPIRE model in \cite{backe2022empire}, Nordpool \cite{nordpoolcap}, Fingrid \cite{fingridcap} and \cite{van2018power}. These capacities are based on the net transfer capacity, i.e., the maximum capacity after taking account for `technical uncertainties on future network conditions' \cite{nordpoolcap}. In practice, the capacity at a given point in time frequently needs to be limited below the net transfer capacity by TSOs for technical reasons. However, we do not model the factors causing such technical reasons explicitly (e.g. detailed characteristics of the national grids) and since capacity restrictions are the key decision variables in our model, we use the net transfer capacities.

Generation capacities of dispatchable generators for each zone are based on annual data from ENTSO-E \cite{entsoe_capacity}, except for the Swedish zones, which are taken from the database of \cite{vanbeesten2022welfare}. We distinguish between the following six types of dispatchable generation: hydropower, nuclear power, combined-cycle gas turbines (CCGT), peaking gas turbines, coal-fired plants an lignite plants. ENTSO-E does not distinguish between CCGT and peaking gas plants but reports single figures for natural gas plants. The model assumes a $2/3$ vs. $1/3$ split between CCGT and peaking gas plants, respectively. To account for the fact that generators are not always available (due to (un)planned maintenance, for instance), we multiply the ENTSO-E capacities by the following availability factors: 85\% for coal and lignite plants, 92\% for nuclear plants, 95\% for CCGT and gas plants. These availability factors are based on \cite{forbes_availability}. For hydropower plants, the model contains an additional limit on the total amount of power that can be produced during the planning period (one week). This production limit is calculated by aggregating the historical production over all periods in the corresponding week. That way, the total amount of production remains the same as in the historical sample, while the distribution over the hours in the planning horizon may be changed by the model.

The marginal costs of dispatchable generation are determined in the following manner. For hydropower plants, we assume that the marginal costs are zero. For nuclear power plants, in line with \cite{vanbeesten2022welfare}, we assume fixed marginal costs of \euro15/MWh. For coal, lignite and natural gas plants, the marginal costs consists of fuel costs and CO$_2$ costs. We discuss these in turn by fuel type. Regarding fuel costs, for CCGT plants and gas peaking plants, we assume that gas is converted to electricity with an efficiency of 55\% and 39\%, respectively (based on IEA \cite{IEAefficiency}). We use the daily day-ahead TTF price as reported by Refinitiv Eikon as proxy for the input price of gas.\footnote{For gas and coal plants, the model assumes that generators offer their bids in the day-ahead market, implying that the input prices that were established on yesterdays day-ahead market for inputs are relevant for the electricity market of today.} For coal and lignite plants, we respectively assume electrical efficiencies of 39\% and 38\% (based on Eurostat \cite{eurostat2021lignite}). For coal plants, the coal input price is proxied by the daily month-ahead ARA (Amsterdam Rotterdam Antwerp) API2 CIF coal price, as reported by Refinitiv Eikon. For lignite plants, because there does not exist a liquid market for lignite, we assume fixed fuel costs of \euro10/MWh \cite{oei2016decarbonizing}. Regarding CO$_2$ costs, gas, coal and lignite plants have to buy EU ETS permits for their emissions. These costs are determined by the CO$_2$ intensity of the respective fuel and the CO$_2$ permit price. Based on \cite{volkerquaschningCO2}, we assume the following CO$_2$ intensities (in tCO$_2$ per MWh of fuel input): 0.359 for coal, 0.364 for lignite and for 0.201 natural gas. The EU ETS price is based on the daily spot EU EUA price as reported by Refinitiv Eikon. We point out that our assumptions imply that the marginal costs for a given type of dispatchable generation do not differ between regions in the model.

The hourly production from variable renewable sources (wind and solar) are based on actual historical production in each zone, which is extracted from ENTSO-E \cite{entsoe_generation}. This production is taken as exogenously given. We assume a corresponding marginal cost of zero.

Linear demand curves are constructed using hourly electricity consumption and day-ahead electricity price data in each zone. Prices are extracted from ENTSO-E,\cite{entsoe_prices},\footnote{Missing values for electricity prices have been manipulated in two ways: (i) in case of six or fewer consecutive missing observations, the observations have been replaced by the average of the last and next known observation; and (ii) in case of more than six consecutive missing prices, the observations have been replaced with the price for the same hour of the zone with the highest correlation coefficient in the sample (e.g. missing prices in SE2 are replaced by the prices from SE2).} while consumption is extracted from ENTSO-E \cite{entsoe_load} for all countries except for the Swedish zones, which comes from the Swedish TSO \cite{swedenTSOmimer}.\footnote{Missing value for consumption have been replaced by one of the following: (a) consumption in the zone on the same hour on the previous day, (b) consumption in the zone on the same hour on the next day, or (c) consumption in the zone on the same hour on the same day in the next week. Option (a) is used when available; if not available, (b) is used; if both (a) and (b) are unavailable, (c) is used.} Assuming that the inverse demand curve is written as $\pi = a d + b$, and using a fixed price elasticity of demand of $\varepsilon = -0.05$ (which is in line with estimates from, e.g., \cite{matar2018households}), the parameters of the inverse demand curve are calculated as
\begin{align*}
    a = \frac{1}{\varepsilon} \frac{|P|}{D}, \qquad b = \left(1 - \frac{1}{\varepsilon}\right) |P|,
\end{align*}
with $P$ and $D$ the historical price and consumption for a particular hour, respectively. We use the absolute value $|P|$ to account for historical prices with negative prices. This method ensures that demand is downward sloping with an elasticity close to $-0.05$ in cases when model outcomes are close to historical outcomes.

Most parameters described above vary over time and thus over our experiments. The only exceptions are the physical capacities of interconnectors and the availability factors for dispatchable generation, which are both fixed. Among the sample-dependent parameters, dispatchable generation capacities vary annually (i.e., for every week the capacity from the corresponding year is taken), marginal costs of dispatchable generation vary daily, and all other parameters vary hourly.

\end{appendices}

\bibliography{references, new_references}

\begin{thebibliography}{40}
\expandafter\ifx\csname natexlab\endcsname\relax\def\natexlab#1{#1}\fi
\providecommand{\url}[1]{\texttt{#1}}
\providecommand{\href}[2]{#2}
\providecommand{\path}[1]{#1}
\providecommand{\DOIprefix}{doi:}
\providecommand{\ArXivprefix}{arXiv:}
\providecommand{\URLprefix}{URL: }
\providecommand{\Pubmedprefix}{pmid:}
\providecommand{\doi}[1]{\href{http://dx.doi.org/#1}{\path{#1}}}
\providecommand{\Pubmed}[1]{\href{pmid:#1}{\path{#1}}}
\providecommand{\bibinfo}[2]{#2}
\ifx\xfnm\relax \def\xfnm[#1]{\unskip,\space#1}\fi
\bibitem[{Beugelsdijk et~al.(2013)Beugelsdijk, Brakman, Garretsen, and van
  Marrewijk}]{beugelsdijk2013international}
\bibinfo{author}{S.~Beugelsdijk}, \bibinfo{author}{S.~Brakman},
  \bibinfo{author}{H.~Garretsen}, \bibinfo{author}{C.~van Marrewijk},
  \bibinfo{title}{{International economics and business: Nations and firms in
  the global economy}}, \bibinfo{edition}{2} ed., \bibinfo{publisher}{Cambridge
  University Press}, \bibinfo{year}{2013}.
\bibitem[{Moradi-Sepahvand and Amraee(2021)}]{moradi2021integrated}
\bibinfo{author}{M.~Moradi-Sepahvand}, \bibinfo{author}{T.~Amraee},
\newblock \bibinfo{title}{Integrated expansion planning of electric energy
  generation, transmission, and storage for handling high shares of wind and
  solar power generation},
\newblock \bibinfo{journal}{Applied Energy} \bibinfo{volume}{298}
  (\bibinfo{year}{2021}) \bibinfo{pages}{117137}.
\bibitem[{{Horn, Henrik and Tanger{\aa}s, Thomas
  P}(2021)}]{horn2021national_new}
\bibinfo{author}{{Horn, Henrik and Tanger{\aa}s, Thomas P}},
\newblock \bibinfo{title}{{National Transmission System Operators in an
  International Electricity Market}},
\newblock \bibinfo{journal}{IFN Working Paper} \bibinfo{volume}{No. 1394}
  (\bibinfo{year}{2021}).
\bibitem[{Glachant and Pignon(2005)}]{glachant2005nordic}
\bibinfo{author}{J.-M. Glachant}, \bibinfo{author}{V.~Pignon},
\newblock \bibinfo{title}{{Nordic congestion's arrangement as a model for
  Europe? Physical constraints vs. economic incentives}},
\newblock \bibinfo{journal}{Utilities Policy} \bibinfo{volume}{13}
  (\bibinfo{year}{2005}) \bibinfo{pages}{153--162}.
\bibitem[{Denmark(2018)}]{winddenmark2018}
\bibinfo{author}{D.~W. E. A. .~W. Denmark}, \bibinfo{title}{Complaint to
  {Swedish energy markets Inspectorate} on curtailment of the two
  {Danish-Swedish} interconnectors}, \bibinfo{year}{2018}. \URLprefix
  \url{https://tinyurl.com/winddenmarkcomplaint}, \bibinfo{note}{last accessed
  on September 30, 2022}.
\bibitem[{Statnett(2021)}]{Statnett2021letter}
\bibinfo{author}{Statnett}, \bibinfo{title}{{Need for a balanced exchange with
  Sweden}}, \bibinfo{year}{2021}. \URLprefix
  \url{https://www.statnett.no/en/about-statnett/news-and-press-releases/news-archive-2021/need-for-a-balanced-exchange-with-sweden/},
  \bibinfo{note}{last accessed on September 30, 2022}.
\bibitem[{e24(2021)}]{e24NorwayConcerned}
\bibinfo{author}{e24}, \bibinfo{title}{Svenskene holder igjen p{\aa} strømmen:
  – vi er bekymret}, \bibinfo{year}{2021}. \URLprefix
  \url{https://e24.no/norsk-oekonomi/i/Kzaz75/svenskene-holder-igjen-paa-stroemmen-vi-er-bekymret},
  \bibinfo{note}{last accessed on September 30, 2022}.
\bibitem[{Hobbs et~al.(2005)Hobbs, Rijkers, and Boots}]{hobbs2005more}
\bibinfo{author}{B.~F. Hobbs}, \bibinfo{author}{F.~A. Rijkers},
  \bibinfo{author}{M.~G. Boots},
\newblock \bibinfo{title}{{The more cooperation, the more competition? A
  Cournot analysis of the benefits of electric market coupling}},
\newblock \bibinfo{journal}{The Energy Journal} \bibinfo{volume}{26}
  (\bibinfo{year}{2005}).
\bibitem[{Pellini(2012)}]{pellini2012measuring}
\bibinfo{author}{E.~Pellini},
\newblock \bibinfo{title}{{Measuring the impact of market coupling on the
  Italian electricity market}},
\newblock \bibinfo{journal}{Energy Policy} \bibinfo{volume}{48}
  (\bibinfo{year}{2012}) \bibinfo{pages}{322--333}.
\bibitem[{Ochoa and {van Ackere}(2015)}]{OCHOA2015522}
\bibinfo{author}{C.~Ochoa}, \bibinfo{author}{A.~{van Ackere}},
\newblock \bibinfo{title}{Winners and losers of market coupling},
\newblock \bibinfo{journal}{Energy} \bibinfo{volume}{80} (\bibinfo{year}{2015})
  \bibinfo{pages}{522--534}.
\bibitem[{Van~den Bergh et~al.(2016)Van~den Bergh, Boury, and
  Delarue}]{van2016flow}
\bibinfo{author}{K.~Van~den Bergh}, \bibinfo{author}{J.~Boury},
  \bibinfo{author}{E.~Delarue},
\newblock \bibinfo{title}{{The flow-based market coupling in central western
  europe: Concepts and definitions}},
\newblock \bibinfo{journal}{The Electricity Journal} \bibinfo{volume}{29}
  (\bibinfo{year}{2016}) \bibinfo{pages}{24--29}.
\bibitem[{Kristiansen(2020)}]{kristiansen2020flow}
\bibinfo{author}{T.~Kristiansen},
\newblock \bibinfo{title}{{The flow based market coupling arrangement in
  Europe: Implications for traders}},
\newblock \bibinfo{journal}{Energy Strategy Reviews} \bibinfo{volume}{27}
  (\bibinfo{year}{2020}) \bibinfo{pages}{100444}.
\bibitem[{Lumbreras and Ramos(2016)}]{lumbreras2016new}
\bibinfo{author}{S.~Lumbreras}, \bibinfo{author}{A.~Ramos},
\newblock \bibinfo{title}{The new challenges to transmission expansion
  planning. survey of recent practice and literature review},
\newblock \bibinfo{journal}{Electric Power Systems Research}
  \bibinfo{volume}{134} (\bibinfo{year}{2016}) \bibinfo{pages}{19--29}.
\bibitem[{Mahdavi et~al.(2018)Mahdavi, Antunez, Ajalli, and
  Romero}]{mahdavi2018transmission}
\bibinfo{author}{M.~Mahdavi}, \bibinfo{author}{C.~S. Antunez},
  \bibinfo{author}{M.~Ajalli}, \bibinfo{author}{R.~Romero},
\newblock \bibinfo{title}{Transmission expansion planning: Literature review
  and classification},
\newblock \bibinfo{journal}{IEEE Systems Journal} \bibinfo{volume}{13}
  (\bibinfo{year}{2018}) \bibinfo{pages}{3129--3140}.
\bibitem[{Gomes and Saraiva(2019)}]{gomes2019state}
\bibinfo{author}{P.~V. Gomes}, \bibinfo{author}{J.~T. Saraiva},
\newblock \bibinfo{title}{State-of-the-art of transmission expansion planning:
  A survey from restructuring to renewable and distributed electricity
  markets},
\newblock \bibinfo{journal}{International Journal of Electrical Power \& Energy
  Systems} \bibinfo{volume}{111} (\bibinfo{year}{2019})
  \bibinfo{pages}{411--424}.
\bibitem[{Churkin et~al.(2021)Churkin, Bialek, Pozo, Sauma, and
  Korgin}]{churkin2021review}
\bibinfo{author}{A.~Churkin}, \bibinfo{author}{J.~Bialek},
  \bibinfo{author}{D.~Pozo}, \bibinfo{author}{E.~Sauma},
  \bibinfo{author}{N.~Korgin},
\newblock \bibinfo{title}{Review of cooperative game theory applications in
  power system expansion planning},
\newblock \bibinfo{journal}{Renewable and Sustainable Energy Reviews}
  \bibinfo{volume}{145} (\bibinfo{year}{2021}) \bibinfo{pages}{111056}.
\bibitem[{ACER(2022)}]{ACER2022report}
\bibinfo{author}{ACER}, \bibinfo{title}{{Report on the result of monitoring the
  margin available for cross-zonal electricity trade in the EU in 2021}},
  \bibinfo{year}{2022}. \URLprefix
  \url{acer.europa.eu/sites/default/files/documents/Publications/ACER\%20MACZT\%20Report\%202021.pdf},
  \bibinfo{note}{last accessed on September 30, 2022}.
\bibitem[{Huang et~al.(2016)Huang, De~Vos, Van~Hertem, Olmos, Rivier, Van~der
  Welle, and Sijm}]{huang2016mind}
\bibinfo{author}{D.~Huang}, \bibinfo{author}{K.~De~Vos},
  \bibinfo{author}{D.~Van~Hertem}, \bibinfo{author}{L.~Olmos},
  \bibinfo{author}{M.~Rivier}, \bibinfo{author}{A.~Van~der Welle},
  \bibinfo{author}{J.~Sijm},
\newblock \bibinfo{title}{Mind the gap: Challenges and policy options for
  cross-border transmission network investments},
\newblock in: \bibinfo{booktitle}{2016 13th International Conference on the
  European Energy Market (EEM)}, \bibinfo{organization}{IEEE},
  \bibinfo{year}{2016}, pp. \bibinfo{pages}{1--6}.
\bibitem[{van Beesten et~al.(2022)van Beesten, Ådnanes, Linde, Pisciella, and
  Tomasgard}]{vanbeesten2022welfare}
\bibinfo{author}{E.~R. van Beesten}, \bibinfo{author}{O.~K. Ådnanes},
  \bibinfo{author}{H.~M. Linde}, \bibinfo{author}{P.~Pisciella},
  \bibinfo{author}{A.~Tomasgard},
\newblock \bibinfo{title}{Welfare compensation in international transmission
  expansion planning under uncertainty},
\newblock \bibinfo{journal}{arXiv preprint arXiv:2205.05978}
  (\bibinfo{year}{2022}).
\bibitem[{Gabriel et~al.(2012)Gabriel, Conejo, Fuller, Hobbs, and
  Ruiz}]{gabriel2012}
\bibinfo{author}{S.~A. Gabriel}, \bibinfo{author}{A.~J. Conejo},
  \bibinfo{author}{J.~D. Fuller}, \bibinfo{author}{B.~F. Hobbs},
  \bibinfo{author}{C.~Ruiz}, \bibinfo{title}{Complementarity modeling in energy
  markets}, \bibinfo{publisher}{Springer Science \& Business Media},
  \bibinfo{year}{2012}.
\bibitem[{EFET(2022)}]{efet2022insight}
\bibinfo{author}{EFET}, \bibinfo{title}{{EFET} insight into accessing
  cross-border transmission capacity in electricity}, \bibinfo{year}{2022}.
  \URLprefix
  \url{https://efet.org/files/documents/20220406_EFET_Insight_7_Transmission_Capacity.pdf},
  \bibinfo{note}{last accessed on September 30, 2022}.
\bibitem[{{NEMO Committee}(2020)}]{euphemia}
\bibinfo{author}{{NEMO Committee}}, \bibinfo{title}{{EUPHEMIA Public
  Description}}, \bibinfo{year}{2020}. \URLprefix
  \url{https://www.nordpoolgroup.com/globalassets/download-center/single-day-ahead-coupling/euphemia-public-description.pdf},
  \bibinfo{note}{last accessed on October 11, 2022}.
\bibitem[{Samuelson(1952)}]{samuelson1952}
\bibinfo{author}{P.~A. Samuelson},
\newblock \bibinfo{title}{Spatial price equilibrium and linear programming},
\newblock \bibinfo{journal}{The American economic review} \bibinfo{volume}{42}
  (\bibinfo{year}{1952}) \bibinfo{pages}{283--303}.
\bibitem[{Dunning et~al.(2017)Dunning, Huchette, and
  Lubin}]{DunningHuchetteLubin2017}
\bibinfo{author}{I.~Dunning}, \bibinfo{author}{J.~Huchette},
  \bibinfo{author}{M.~Lubin},
\newblock \bibinfo{title}{Jump: A modeling language for mathematical
  optimization},
\newblock \bibinfo{journal}{SIAM Review} \bibinfo{volume}{59}
  (\bibinfo{year}{2017}) \bibinfo{pages}{295--320}.
\bibitem[{{Gurobi Optimization, LLC}(2022)}]{gurobi_new}
\bibinfo{author}{{Gurobi Optimization, LLC}}, \bibinfo{title}{{Gurobi Optimizer
  Reference Manual}}, \bibinfo{year}{2022}. \URLprefix
  \url{https://www.gurobi.com}, \bibinfo{note}{last accessed on September 30,
  2022}.
\bibitem[{Backe et~al.(2022)Backe, Skar, del Granado, Turgut, and
  Tomasgard}]{backe2022empire}
\bibinfo{author}{S.~Backe}, \bibinfo{author}{C.~Skar}, \bibinfo{author}{P.~C.
  del Granado}, \bibinfo{author}{O.~Turgut}, \bibinfo{author}{A.~Tomasgard},
\newblock \bibinfo{title}{Empire: An open-source model based on multi-horizon
  programming for energy transition analyses},
\newblock \bibinfo{journal}{SoftwareX} \bibinfo{volume}{17}
  (\bibinfo{year}{2022}) \bibinfo{pages}{100877}.
\bibitem[{{Fingrid}(2020)}]{nordpoolcap}
\bibinfo{author}{{Fingrid}}, \bibinfo{title}{Principles for determining the net
  transfer capacities in the nordic power market}, \bibinfo{year}{2020}.
  \URLprefix
  \url{https://www.nordpoolgroup.com/4aad73/globalassets/download-center/tso/principles-for-determining-the-transfer-capacities_2020-09-22.pdf},
  \bibinfo{note}{last accessed on September 30, 2022}.
\bibitem[{{Fingrid}(2017)}]{fingridcap}
\bibinfo{author}{{Fingrid}}, \bibinfo{title}{Notice on capacity},
  \bibinfo{year}{2017}. \URLprefix
  \url{https://www.fingrid.fi/en/electricity-market/rajajohto-informaatio/notice-on-capacity/},
  \bibinfo{note}{last accessed on September 30, 2022}.
\bibitem[{Van~Leeuwen and Mulder(2018)}]{van2018power}
\bibinfo{author}{C.~Van~Leeuwen}, \bibinfo{author}{M.~Mulder},
\newblock \bibinfo{title}{Power-to-gas in electricity markets dominated by
  renewables},
\newblock \bibinfo{journal}{Applied Energy} \bibinfo{volume}{232}
  (\bibinfo{year}{2018}) \bibinfo{pages}{258--272}.
\bibitem[{{ENTSO-E}(2022)}]{entsoe_capacity}
\bibinfo{author}{{ENTSO-E}}, \bibinfo{title}{Installed capacity per production
  type}, \bibinfo{year}{2022}. \URLprefix
  \url{https://transparency.entsoe.eu/generation/r2/installedGenerationCapacityAggregation/show},
  \bibinfo{note}{last accessed on September 30, 2022}.
\bibitem[{{Forbes}(2017)}]{forbes_availability}
\bibinfo{author}{{Forbes}}, \bibinfo{title}{Does `fuel on hand' make coal and
  nuclear power plants more valuable}, \bibinfo{year}{2017}. \URLprefix
  \url{https://www.forbes.com/sites/amorylovins/2017/05/01/does-fuel-on-hand-make-coal-and-nuclear-power-plants-more-valuable/},
  \bibinfo{note}{last accessed on September 30, 2022}.
\bibitem[{{IEA}(2020)}]{IEAefficiency}
\bibinfo{author}{{IEA}}, \bibinfo{title}{{European Union marginal coal- and
  gas-fired power generation costs, 2018-2021}}, \bibinfo{year}{2020}.
  \URLprefix
  \url{https://www.iea.org/data-and-statistics/charts/european-union-marginal-coal-and-gas-fired-power-generation-costs-2018-2021},
  \bibinfo{note}{last accessed on September 30, 2022}.
\bibitem[{Eurostat(2021)}]{eurostat2021lignite}
\bibinfo{author}{Eurostat}, \bibinfo{title}{{Production of lignite in the EU -
  statistics}}, \bibinfo{year}{2021}. \URLprefix
  \url{https://ec.europa.eu/eurostat/statistics-explained/index.php?title=Production_of_lignite_in_the_EU_-_statistics#What_technology_is_used_to_transform_lignite_and_how_efficient_is_it.3F},
  \bibinfo{note}{last accessed on September 30, 2022}.
\bibitem[{Oei(2016)}]{oei2016decarbonizing}
\bibinfo{author}{P.-Y. C.~R. Oei}, \bibinfo{title}{Decarbonizing the European
  Electricity Sector: Modeling and Policy Analysis for Electricity and CO$_2$
  Infrastructure Networks [Dissertation]}, \bibinfo{publisher}{Technische
  Universitaet Berlin (Germany)}, \bibinfo{year}{2016}.
\bibitem[{{Volker-Quaschning}(2021)}]{volkerquaschningCO2}
\bibinfo{author}{{Volker-Quaschning}}, \bibinfo{title}{{Specific Carbon Dioxide
  Emissions of Various Fuels}}, \bibinfo{year}{2021}. \URLprefix
  \url{https://www.volker-quaschning.de/datserv/CO2-spez/index_e.php},
  \bibinfo{note}{last accessed on September 30, 2022}.
\bibitem[{{ENTSO-E}(2022{\natexlab{a}})}]{entsoe_generation}
\bibinfo{author}{{ENTSO-E}}, \bibinfo{title}{Actual generation per production
  type}, \bibinfo{year}{2022}{\natexlab{a}}. \URLprefix
  \url{https://transparency.entsoe.eu/generation/r2/actualGenerationPerProductionType/show},
  \bibinfo{note}{last accessed on September 30, 2022}.
\bibitem[{{ENTSO-E}(2022{\natexlab{b}})}]{entsoe_prices}
\bibinfo{author}{{ENTSO-E}}, \bibinfo{title}{Day-ahead prices},
  \bibinfo{year}{2022}{\natexlab{b}}. \URLprefix
  \url{https://transparency.entsoe.eu/transmission-domain/r2/dayAheadPrices/show},
  \bibinfo{note}{last accessed on September 30, 2022}.
\bibitem[{{ENTSO-E}(2022{\natexlab{c}})}]{entsoe_load}
\bibinfo{author}{{ENTSO-E}}, \bibinfo{title}{Total load - day ahead / actual},
  \bibinfo{year}{2022}{\natexlab{c}}. \URLprefix
  \url{https://transparency.entsoe.eu/load-domain/r2/totalLoadR2/show},
  \bibinfo{note}{last accessed on September 30, 2022}.
\bibitem[{Mimer(2022)}]{swedenTSOmimer}
\bibinfo{author}{Mimer}, \bibinfo{title}{F\"{o}rbrukningsstatistik},
  \bibinfo{year}{2022}. \URLprefix
  \url{https://mimer.svk.se/ProductionConsumption/ConsumptionIndex},
  \bibinfo{note}{last accessed on September 30, 2022}.
\bibitem[{Matar(2018)}]{matar2018households}
\bibinfo{author}{W.~Matar},
\newblock \bibinfo{title}{Households' response to changes in electricity
  pricing schemes: bridging microeconomic and engineering principles},
\newblock \bibinfo{journal}{Energy Economics} \bibinfo{volume}{75}
  (\bibinfo{year}{2018}) \bibinfo{pages}{300--308}.

\end{thebibliography}

\end{document}